\documentclass{article}%
\usepackage{amssymb}
\usepackage{amsfonts}
\usepackage{amsmath}%
\usepackage{graphicx}
\providecommand{\U}[1]{\protect\rule{.1in}{.1in}}
\newtheorem{theorem}{Theorem}

\newtheorem{corollary}[theorem]{Corollary}

\newtheorem{definition}[theorem]{Definition}

\newtheorem{proposition}[theorem]{Proposition}
\newtheorem{remark}[theorem]{Remark}

\newenvironment{proof}[1][Proof]{\noindent\textbf{#1.} }{\ \rule{0.5em}{0.5em}}
\begin{document}

\title{Non-Hamiltonian systems separable by Hamilton-Jacobi method}
\author{Krzysztof Marciniak$^{\ast}$\\Department of Science and Technology \\Campus Norrk\"{o}ping, Link\"{o}ping University\\601-74 Norrk\"{o}ping, Sweden\\krzma@itn.liu.se
\and Maciej B\l aszak\thanks{Partially supported by Swedish Research Council grant
no. VR 2006-7359 and MNiSW research grant N202}\\Institute of Physics, A. Mickiewicz University\\Umultowska 85, 61-614 Pozna\'{n}, Poland\\blaszakm@amu.edu.pl}
\maketitle

\begin{abstract}
We show that with every separable calssical St\"{a}ckel system of Benenti type
on a Riemannian space one can associate, by a proper deformation of the metric
tensor, a multi-parameter family of non-Hamiltonian systems on the same space,
sharing the same trajectories and related to the seed system by appropriate
reciprocal transformations. These system are known as bi-cofactor systems and
are integrable in quadratures as the seed Hamiltonian system is. We show that
with each class of bi-cofactor systems a pair of separation curves can be
related. We also investigate conditions under which a given flat bi-cofactor
system can be deformed to a family of geodesically equivalent flat bi-cofactor systems.

\end{abstract}


\section{Introduction}

A significant progress in the geometric separability theory for classical
Hamiltonian systems separable by Hamilton-Jacobi method has been achieved in
recent years (see for example \cite{sk1} - \cite{m3}). Among other things a
new class of non-Hamiltonian Newton systems was introduced \cite{skh},
\cite{h}. These systems were shown to have very interesting geometric
properties when considered as systems on Riemann spaces \cite{sc1},\cite{sc2}
(see also \cite{c1}). In \cite{km} we showed that they can be separated by the
Hamilton-Jacobi method after certain reparametrization of the evolution
parameter (see also \cite{StefanClaes}). Originally these systems were called
quasi-Lagrangian systems. In the present literature they are called
bi-cofactor systems or cofactor-pair systems. In \cite{be1} it was further
shown that each bi-cofactor system is \emph{geodesically equivalent} (in the
classical sense of Levi-Civita \cite{Levi-Civita}) to some separable
Lagrangian system which means that it has the same trajectories on the
underlying configuration manifold as the Lagrangian system only traversed with
a different speed and moreover that the metric tensors associated with both
systems are equivalent i.e. have the same geodesics (considered as
unparametrized curves). In the same paper one can also find a thorough
geometric theory of bi-cofactor systems on an arbitrary pseudoriemannian space.

In the present paper we demonstrate on the level of differential equations the
geodesic equivalence properties of cofactor and bi-cofactor systems expressed
by an appropriate class of reciprocal transformations. We clarify and
systematize their bi-quasihamiltonian formulation on the phase space. We show
explicitly that a bi-cofactor system is geodesically equivalent to two
different separable Hamiltonian systems of Benenti type and we show explicitly
the transformation between all geometric structures associated with these two
Benenti systems and the original bi-cofactor system. We further demonstrate
that with each bi-cofactor system one can relate two different separation
curves and we find a map between these curves. From this point of view we
therefore show that with each pair of separation curves that are related
through the above mentioned map we can associate a whole class of geodesically
equivalent bi-cofactor systems. Every such class contains at least two
separable Hamiltonian systems and on the phase space all the members of a
given class are related by a composition of an appropriate non-canonical
transformation and a reciprocal transformation. Further, we investigate
geodesically equivalent families of flat (in the sense of the underlying
metric tensor) cofactor systems and find a sufficient condition for a so
called $J$-tensor to generate from any given flat bi-cofactor system a
multi-parameter family of flat bi-cofactor systems. Finally, we illustrate our
considerations by presenting a thorough example of the class of separable
bi-cofactor systems geodesically equivalent to the Henon-Heiles system and
then specify this example to the flat case.

\section{Cofactor systems}

Let us consider the following Newton system%
\begin{equation}
\frac{d^{2}q^{i}}{dt^{2}}+\Gamma_{jk}^{i}\frac{dq^{j}}{dt}\frac{dq^{k}}%
{dt}=F^{i},\text{ \ \ \ \ }i=1,\ldots,n\label{1}%
\end{equation}
where $q^{i}$ are some coordinates on an $n$-dimensional pseudo-Riemannian
manifold $Q$ endowed with a metric tensor $g=(g_{ij})$ and where $F=(F^{i})$
is a vector field on $Q$ representing the force which we assume time- and
velocity-independent. Here and in what follows we use the Einstein summation
convention if not stated otherwise. The functions $\Gamma_{jk}^{i}$ are the
Christoffel symbols of the Levi-Civita connection associated with the metric
tensor $g$ and if all $\Gamma_{jk}^{i}$ are zero we call the system (\ref{1})
a \emph{flat Newton system}. In case that $F=0$ (\ref{1}) is the equation of
geodesic motion on $Q$ and the variable $t$ becomes an affine parameter of
geodesic lines.

If the force $F$ is conservative (potential) i.e. if
\begin{equation}
F=-\nabla V=-GdV,\label{3}%
\end{equation}
where $G=g^{-1}$ is the contravariant form of the metric tensor $g$ and where
$V=V(q)$ is a potential function, then (\ref{1}) is equivalent to the
Lagrangian system
\begin{equation}
\frac{d}{dt}\frac{\partial\mathcal{L}}{\partial v^{i}}-\frac{\partial
\mathcal{L}}{\partial q^{i}}=0,\text{ \ }v^{i}=\frac{d}{dt}q^{i}%
,\ \ \ \text{\ }i=1,\ldots,n\label{2}%
\end{equation}
on the tangent bundle $TQ$ endowed with coordinates $(q,v)=(q^{1},\ldots
q^{n},v^{1},\ldots,v^{n}),$ where $\mathcal{L}=\frac{1}{2}g_{ij}(q)v^{i}%
v^{j}-V(q)$ is a Lagrangian of the system. By the Legendre map $p_{i}%
=g_{ij}v^{j}$ the system (\ref{2}) is transformed to the Hamiltonian dynamical
system%
\begin{equation}
\frac{d}{dt}\left(
\begin{array}
[c]{c}%
q\\
p
\end{array}
\right)  =\left(
\begin{array}
[c]{cc}%
0 & I\\
-I & 0
\end{array}
\right)  \left(
\begin{array}
[c]{c}%
\frac{\partial H}{\partial q}\\
\frac{\partial H}{\partial p}%
\end{array}
\right)  =\Pi_{c}\,dH\label{ham}%
\end{equation}
on the cotangent bundle $T^{\ast}Q$ endowed with coordinates $(q,p)=(q^{i}%
,p_{j})$ where $H=\frac{1}{2}G^{ij}(q)p_{i}p_{j}+V(q)$ is the Hamiltonian of
the system, $\Pi_{c}$ is the canonical Poisson tensor and $dH $ is the
differential of $H$.

We will now remind the notion of a $J$-tensor.

\begin{definition}
A $(1,1)$-tensor $\mathbf{J}=(J_{j}^{i})$ on $Q$ is called a $J$-tensor
associated with the metric $g$ or $G=g^{-1}$ (we often write that $\mathbf{J}
$ is a $J_{G}$-tensor when emphasizing the underlying metric) if its
contravariant form $J^{ij}=J_{k}^{i}G^{kj}$ is a symmetric $(2,0)$-tensor and
if $\mathbf{J}$ itself satisfies the following \emph{characteristic equation}%
\begin{equation}
\nabla_{h}J_{j}^{i}=\left(  \alpha_{j}\delta_{h}^{i}+\alpha^{i}g_{jh}\right)
\label{charJ}%
\end{equation}
where $\nabla_{h}$ is the covariant derivative associated with the metric $g $
and where $\alpha_{i}$ is some $1$-form.
\end{definition}

From (\ref{charJ}) it follows that the Nijenhuis torsion of $\mathbf{J}$
vanishes:%
\[
J_{\left[  i\right.  }^{h}\nabla_{\left\vert h\right\vert }J_{\left.
j\right]  }^{k}-J_{l}^{k}\nabla_{\left[  i\right.  }J_{\left.  j\right]  }%
^{l}=0
\]
(the square brackets denote skew-symmetric permutations of indices $i,j;$ the
index $h$ is not permuted) and that $\mathbf{J}$ is a conformal Killing tensor
of trace type which means that $J_{ij}=J_{i}^{k}g_{kj}$ satisfies the relation
$\nabla_{\left(  h\right.  }J_{\left.  ij\right)  }=\alpha_{\left(  h\right.
}g_{\left.  ij\right)  }$ with $\alpha_{i}=\partial_{i}\operatorname*{tr}%
\mathbf{J}$ (the brackets denote symmetric permutations of indices $h,i,j$).

\begin{remark}
\label{Jspace}All $J$-tensors of a given metric tensor $g$ constitute an
$\mathbf{R}$-linear vector space of dimension less or equal to $\frac{1}%
{2}(n+1)(n+2)$. This space attains its maximum dimension for metrics of
constant curvature. In case the metric $g$ is pseudoeuclidean so that
$g=\operatorname*{diag}(\varepsilon_{1},\ldots,\varepsilon_{n})$ with
$\varepsilon_{i}=\pm1$ in its Cartesian coordinates, the general form of
$\mathbf{J}$ in these coordinates is \cite{be1}%
\begin{equation}
J^{ij}=mq^{i}q^{j}+\beta^{i}q^{j}+\beta^{j}q^{i}+\gamma^{ij}\label{og}%
\end{equation}
where $m,\beta^{i}$ and $\gamma^{ij}=\gamma^{ji}$ are $\frac{1}{2}(n+1)(n+2)$
independent constants and where $J^{ij}=J_{k}^{i}G^{kj}$ is the contravariant
form of $\mathbf{J}$.
\end{remark}

If a $J$-tensor $\mathbf{J}$ has $n$ real and simple eigenvalues then it is
called $L$-tensor and its signed eigenvalues $(\lambda^{1},\ldots,\lambda
^{n})$ given by $\det\left(  \mathbf{J}+\lambda(q)I\right)  =0$ define a
coordinate web on $Q$. Such webs will turn out to be separation webs for our
systems (see below). See \cite{be1} for further details on $J$-tensors and $L
$-tensors.

The system (\ref{1}) is called \emph{cofactor} if the force $F$ has the
following form%
\begin{equation}
F=-\left(  \operatorname*{cof}\mathbf{J}\right)  ^{-1}\nabla V\label{cof}%
\end{equation}
for some $J$-tensor $\mathbf{J}$, where $\operatorname*{cof}\mathbf{J}$ is the
cofactor matrix of $\mathbf{J}$\textbf{\ }(i.e. the transposed matrix of
signed minors of $\mathbf{J}$) so that $\mathbf{J}\operatorname*{cof}%
\mathbf{J=}\left(  \operatorname*{cof}\mathbf{\mathbf{J}}\right)
\mathbf{\mathbf{\,}J=}\left(  \det\mathbf{J}\right)  \mathbf{\,}I$ or in case
that $\mathbf{J}$ is invertible $\operatorname*{cof}\mathbf{J}=\left(
\det\mathbf{J}\right)  \mathbf{\,J}^{-1}$. In the case $\mathbf{J}=I$ the
system (\ref{cof}) becomes Lagrangian (potential).

In our further considerations the notion of equivalent metric tensors will
play an important role. Two metric tensors $G$ and $\overline{G}$ on manifold
$Q$ are said to be \emph{equivalent} if their geodesics locally coincide as
unparametrized curves. As it was shown in \cite{be1}, a metric $G $ admits an
equivalent metric $\overline{G}$ if and only if it admits a nonsingular
$J$-tensor $\mathbf{J}$. In such a case $\overline{G}^{ij}=\sigma J_{k}%
^{i}G^{kj}=\sigma J^{ij}$ or in the matrix form
\[
\overline{G}=\sigma\mathbf{J}G
\]
with $\sigma=\det\mathbf{J}=\frac{dt}{d\overline{t}}$ where $t\,\ $and
$\overline{t}$ are affine parameters associated with the (parametrized)
geodesic of $G$ and $\overline{G}$ respectively. Moreover, $\mathbf{J}^{-1}$
is a $J$-tensor for the new metric $\overline{G}$.

Two dynamical systems $(g,F)$ and $(\overline{g},\overline{F})$ of the form
(\ref{1}) on $Q$ are said to be equivalent if their trajectories coincide up
to a reparametrization of the evolution parameter. Moreover, they are called
\emph{geodesically equivalent} if also metrics $g$ and $\overline{g}$ are
equivalent. As it was proved in \cite{be1} two systems $(g,F)$ and
$(\overline{g},\overline{F})$ are geodesically equivalent if and only if the
metric $g$ admits a nonsingular $J$-tensor $\mathbf{J}$ such that
\[
\overline{G}^{ij}=\sigma J_{k}^{i}G^{kj},\ \ \overline{F}=\sigma
^{2}F\mathbf{,\ \ \ }\ \sigma=\det\mathbf{J}.
\]
Also in this case the evolution parameters $t$ and $\overline{t}$ of the
systems $(g,F)$ and $(\overline{g},\overline{F})$ are related through the
above mentioned reciprocal transformation$\ $%
\[
\frac{dt}{d\overline{t}}=\sigma.
\]
We will now show that every cofactor system belongs to a whole class of
geodesically equivalent cofactor systems.

\begin{theorem}
\label{trans}Consider the cofactor system%
\begin{equation}
\frac{d^{2}q^{i}}{dt^{2}}+\Gamma_{jk}^{i}\frac{dq^{j}}{dt}\frac{dq^{k}}%
{dt}=-\left(  \left(  \operatorname*{cof}\mathbf{J}\right)  ^{-1}\nabla
V\right)  ^{i}\text{, \ \ \ \ }i=1,\ldots,n.\label{coft}%
\end{equation}
Assume that $\mathbf{J}_{1}$ is another $J$-tensor for the metric $G$ and
denote by $G_{1}=\sigma_{1}\mathbf{J}_{1}G$ (with $\sigma_{1}=\det
\mathbf{J}_{1}$) a new metric tensor equivalent to $G$. In a new independent
variable $t_{1}$ defined through the reciprocal transformation%
\[
dt_{1}=\frac{dt}{\sigma_{1}}%
\]
the cofactor system (\ref{coft}) attains the form%
\begin{equation}
\frac{d^{2}q^{i}}{dt_{1}^{\,2}}+(\Gamma^{(1)})_{jk}^{i}\frac{dq^{j}}{dt_{1}%
}\frac{dq^{k}}{dt_{1}}=-\left(  \left[  \operatorname*{cof}\left(
\mathbf{J\,J}_{1}^{-1}\right)  \right]  ^{-1}\nabla^{(1)}V\right)  ^{i}\text{,
\ \ \ \ }i=1,\ldots,n\label{coft1}%
\end{equation}
where $(\Gamma^{(1)})_{jk}^{i}$ are Christoffel symbols of the metric $G_{1} $
and $\nabla^{(1)}=G_{1}d$ is the gradient operator associated with the metric
$G_{1}$.
\end{theorem}

\begin{proof}
Since $dt_{1}=dt/\sigma_{1}$ we have, by the chain rule,
\[
\frac{dq^{i}}{dt}=\frac{1}{\sigma_{1}}\frac{dq^{i}}{dt_{1}}\text{, \ \ }%
\frac{d^{2}q^{i}}{dt^{2}}=\frac{1}{\sigma_{1}^{2}}\text{\ }\frac{d^{2}q^{i}%
}{dt_{1}^{2}}-\frac{1}{\sigma_{1}^{3}}\frac{dq^{i}}{dt_{1}}\frac
{\partial\sigma_{1}}{\partial q_{l}}\frac{dq_{l}}{dt_{1}}.
\]
Moreover (see for example \cite{Schouten}) the Christoffel symbols of $G$ and
$G_{1}$ are related by%
\begin{equation}
\Gamma_{jk}^{i}=(\Gamma^{(1)})_{jk}^{i}+\frac{1}{2\sigma_{1}}\left(
\delta_{j}^{i}\frac{\partial\sigma_{1}}{\partial q_{k}}+\delta_{k}^{i}%
\frac{\partial\sigma_{1}}{\partial q_{j}}\right)  .\label{Gammy}%
\end{equation}
Further
\[
\nabla^{(1)}V=G_{1}dV=\sigma_{1}\mathbf{J}_{1}GdV=\sigma_{1}\mathbf{J}%
_{1}\nabla V,
\]
so that%
\begin{align*}
\left(  \operatorname*{cof}\mathbf{J}\right)  ^{-1}\nabla V  & =\frac
{1}{\sigma_{1}}\left(  \operatorname*{cof}\mathbf{J}\right)  ^{-1}%
\mathbf{J}_{1}^{-1}\nabla^{(1)}V=\frac{1}{\sigma_{1}^{2}}\left(
\operatorname*{cof}\mathbf{J}\right)  ^{-1}\operatorname*{cof}\mathbf{J}%
_{1}\nabla^{(1)}V=\\
& =\frac{1}{\sigma_{1}^{2}}\operatorname*{cof}(\mathbf{J}^{-1}%
)\operatorname*{cof}\mathbf{J}_{1}\nabla^{(1)}V=\frac{1}{\sigma_{1}^{2}%
}\operatorname*{cof}(\mathbf{J}_{1}\mathbf{J}^{-1})\nabla^{(1)}V=\\
& =\frac{1}{\sigma_{1}^{2}}\left(  \operatorname*{cof}\left(  \mathbf{J\,J}%
_{1}^{-1}\right)  ^{-1}\right)  \nabla^{(1)}V.
\end{align*}
Plugging all this into (\ref{coft}) we obtain (\ref{coft1}).
\end{proof}

\begin{remark}
The tensor $\mathbf{J\,J}_{1}^{-1}$ is a $J_{G_{1}}$-tensor i.e. a $J$-tensor
for the metric $G_{1}$. It means that the system (\ref{coft1}) is a cofactor
system geodesically equivalent to (\ref{coft}) with $G_{1}$ as the underlying metric.
\end{remark}

Note that in the particular case $\mathbf{J}_{1}=\mathbf{J}$ the system
(\ref{coft1}) becomes potential%
\begin{equation}
\frac{d^{2}q^{i}}{d\overline{t}^{2}}+\overline{\Gamma}_{jk}^{i}\frac{dq^{j}%
}{d\overline{t}}\frac{dq^{k}}{d\overline{t}}=-\left(  \overline{\nabla
}V\right)  ^{i}\text{, \ \ \ \ }i=1,\ldots,n.\label{cofbar}%
\end{equation}
with the affine parameter
\[
dt_{1}=d\overline{t}=\frac{dt}{\sigma}%
\]
and with $\overline{\Gamma}_{jk}^{i}$ and $\overline{\nabla}=\overline{G}d$
defined by the new metric $\overline{G}=\sigma\mathbf{J}G$ with $\sigma
=\det(\mathbf{J})$. This shows that every cofactor system is geodesically
equivalent (in the sense of the definition above) to a potential system. This
fact yields us a possibility of determining a quasi-hamiltonian formulation
for the cofactor system (\ref{coft}).

\begin{proposition}
\label{HAM}The cofactor system (\ref{coft}) has on $T^{\ast}Q$ the following
quasi-Hamiltonian representation:%
\begin{equation}
\frac{d}{dt}\left(
\begin{array}
[c]{c}%
q\\
p
\end{array}
\right)  =\frac{1}{\sigma}\Pi_{nc}dH\label{hamcof}%
\end{equation}
with the noncanonical Poisson operator
\[
\Pi_{nc}=\left(
\begin{array}
[c]{cc}%
0 & \mathbf{J}\\
-\mathbf{J}^{T} & \Omega
\end{array}
\right)  ,\ \ \ \Omega_{j}^{i}=\left(  \frac{\partial J_{i}^{k}}{\partial
q^{j}}-\frac{\partial J_{j}^{k}}{\partial q^{i}}\right)  p_{k}%
\]
and with the Hamiltonian%
\begin{equation}
H(q,p)=\frac{1}{2}p^{T}(\operatorname*{cof}\text{\thinspace}\mathbf{J}%
)Gp+V(q).\label{Hamcof}%
\end{equation}

\end{proposition}

\begin{proof}
The systems (\ref{coft}) and (\ref{cofbar}) are related by the reciprocal
transformation $d\overline{t}=dt/\sigma$ with $\sigma=\sigma(q)$ yielding that
$dq^{i}/d\overline{t}=\sigma dq^{i}/dt$. Let us thus introduce new variables
on $TQ$:%
\begin{equation}
\overline{q}=q,\text{ }\overline{v}=\sigma v\text{.}\label{zmianav}%
\end{equation}
The Lagrangian of (\ref{cofbar}) written in coordinates $(\overline
{q},\overline{v})$ is:
\[
\overline{\mathcal{L}}=\frac{1}{2}\overline{g}_{ij}(\overline{q})\overline
{v}^{i}\overline{v}^{j}-V(\overline{q})
\]
This Lagrangian defines a new Legendre map from $TQ$ to $T^{\ast}Q$ that is
just the fiberwise isomorphism between $TQ$ and $T^{\ast}Q$ induced by the new
metric $\overline{g}$ i.e. $\overline{p}=\overline{g}\,\overline{v}$. From
$\overline{G}=\sigma\mathbf{J}G$ we have
\[
\overline{g}=\overline{G}^{-1}=\frac{1}{\sigma}g\mathbf{J}^{-1}=\frac
{1}{\sigma}(\mathbf{J}^{T})^{-1}g
\]
so that%
\[
\overline{p}=\overline{g}\,\overline{v}=\frac{1}{\sigma}(\mathbf{J}^{T}%
)^{-1}g\sigma v=(\mathbf{J}^{T})^{-1}gv=(\mathbf{J}^{T})^{-1}p.
\]
Thus, the map (\ref{zmianav}) on $TQ$ induces the following non-canonical map
on $T^{\ast}Q$:%
\begin{equation}
\overline{q}=q\text{, \ }\overline{p}=\left(  \mathbf{J}^{T}\right)
^{-1}p.\label{zmianap}%
\end{equation}
In the coordinates $(\overline{q},\overline{p})$ the system (\ref{cofbar}) has
the following canonical Hamiltonian representation (cf (\ref{ham})):%
\begin{equation}
\frac{d}{d\overline{t}}\left(
\begin{array}
[c]{c}%
\overline{q}\\
\overline{p}%
\end{array}
\right)  =\overline{\Pi}_{c}d\overline{H},\label{hamcofbar}%
\end{equation}
with the usual Hamiltonian $\overline{H}=\frac{1}{2}\overline{p}%
^{T}\,\overline{G\,}\overline{p}+V(\overline{q})$. In order to obtain the
quasi-hamiltonian form (\ref{hamcof}) of (\ref{coft}) it is enough to
transform the system (\ref{hamcofbar}) back to the variables $(q,p,t)$. The
map between these variables is%
\begin{equation}
\overline{q}=q\text{, \ }\overline{p}=\left(  \mathbf{J}^{T}\right)
^{-1}p\text{, \ }d\overline{t}=\frac{dt}{\sigma}\text{.}\label{mapa}%
\end{equation}
or equivalently%
\begin{equation}
q=\overline{q},\,\ p=\mathbf{J}^{T}\overline{p},\text{ \ }dt=\sigma
d\overline{t}\text{.}\label{mapaodwr}%
\end{equation}
Note that this map consists of a "space" part (\ref{cofbar}) that involves
only $(q,p)$ and $(\overline{q},\overline{p})$ variables followed by the
reciprocal transformation (reparametrization of evolution parameter)
$d\overline{t}=dt/\sigma$. By using that $d/d\overline{t}=\sigma d/dt$ (which
generates the factor $1/\sigma$ in (\ref{hamcof})) and after some calculations
that exploit the fact that $\mathbf{J}$ is torsionfree we obtain
(\ref{hamcof}) with $H$ denoting the function $\overline{H}$ written in
$(q,p)$-coordinates. Since
\[
\overline{p}^{T}\,\overline{G\,}\overline{p}=p^{T}\mathbf{J}^{-1}%
\sigma\mathbf{J}G(\mathbf{J}^{T})^{-1}p=p^{T}\sigma\mathbf{J}^{-1}%
Gp=p^{T}\operatorname*{cof}(\mathbf{J})Gp
\]
we get that $H$ is of the form (\ref{Hamcof}).
\end{proof}

\section{Bi-cofactor systems}

The system of Newton equations of the form%
\begin{equation}
\frac{d^{2}q^{i}}{dt^{2}}+\Gamma_{jk}^{i}\frac{dq^{j}}{dt}\frac{dq^{k}}%
{dt}=-\left(  \left(  \operatorname*{cof}\mathbf{J}_{1}\right)  ^{-1}\nabla
V\right)  ^{i}=-\left(  \left(  \operatorname*{cof}\mathbf{J}_{2}\right)
^{-1}\nabla W\right)  ^{i}\label{bicoft}%
\end{equation}
with two independent $J_{G}$-tensors $\mathbf{J}_{1}$ and $\mathbf{J}_{2}$ and
with two different potentials $V$ and $W$ is called a \emph{bi-cofactor
system} on $Q$. It means that the force $F$ has two different cofactor
representations of the form (\ref{cof}). The following is a simple corollary
of Theorem \ref{trans}.

\begin{proposition}
\label{czast3}Assume that the metric $G$ has a third $J$-tensor $\mathbf{J}%
_{3}$ and denote by $G_{3}=\sigma_{3}\mathbf{J}_{3}G$ (with $\sigma_{3}%
=\det\mathbf{J}_{3}$) a new metric tensor equivalent to $G$. In the new
independent variable $t_{3}$ defined through
\begin{equation}
dt_{3}=\frac{dt}{\sigma_{3}}\label{t3}%
\end{equation}
the bi-cofactor system (\ref{bicoft}) attains the form%
\begin{align}
\frac{d^{2}q^{i}}{dt_{3}^{\,2}}+(\Gamma^{(3)})_{jk}^{i}\frac{dq^{j}}{dt_{3}%
}\frac{dq^{k}}{dt_{3}}  & =-\left(  \left[  \operatorname*{cof}\left(
\mathbf{J}_{1}\mathbf{\,J}_{3}^{-1}\right)  \right]  ^{-1}\nabla
^{(3)}V\right)  ^{i}\nonumber\\
& =-\left(  \left[  \operatorname*{cof}\left(  \mathbf{J}_{2}\mathbf{\,J}%
_{3}^{-1}\right)  \right]  ^{-1}\nabla^{(3)}W\right)  ^{i}\label{bicoft3}%
\end{align}
where $(\Gamma^{(3)})_{jk}^{i}$ are Christoffel symbols of the metric $G_{3} $
and $\nabla^{(3)}=G_{3}d$.
\end{proposition}

As before, both the tensor $\mathbf{J}_{1}\mathbf{\,J}_{3}^{-1}$ and
$\mathbf{J}_{2}\mathbf{\,J}_{3}^{-1}$ are $J_{G_{3}}$-tensors so that $G_{3}$
is the underlying metric of the system (\ref{bicoft3}).

In case that $\mathbf{J}_{3}=\mathbf{J}_{1}$ the system (\ref{bicoft3})
attains the potential-cofactor form%

\begin{equation}
\frac{d^{2}q^{i}}{d\overline{t}^{2}}+\overline{\Gamma}_{jk}^{i}\frac{dq^{j}%
}{d\overline{t}}\frac{dq^{k}}{d\overline{t}}=-\left(  \overline{\nabla
}V\right)  ^{i}=-\left(  \left(  \operatorname*{cof}\overline{\mathbf{J}%
}\right)  ^{-1}\overline{\nabla}W\right)  ^{i}\label{bar}%
\end{equation}
with the affine parameter $d\overline{t}=dt/\overline{\sigma}$ and with
$\overline{\mathbf{J}}=\mathbf{J}_{2}\mathbf{\,J}_{1}^{-1}$ being a
$J_{\overline{G}}$-tensor for the new metric $\overline{G}=\overline{\sigma
}\mathbf{J}_{1}G$ with $\overline{\sigma}=\det(\mathbf{J}_{1})=\sigma_{1}$.

If $\mathbf{J}_{3}=\mathbf{J}_{2}$ then the system (\ref{bicoft3}) attains the
cofactor-potential form%

\begin{equation}
\frac{d^{2}q^{i}}{d\widetilde{t}^{2}}+\widetilde{\Gamma}_{jk}^{i}\frac{dq^{j}%
}{d\widetilde{t}}\frac{dq^{k}}{d\widetilde{t}}=-\left(  \left(
\operatorname*{cof}\widetilde{\mathbf{J}}\right)  ^{-1}\widetilde{\nabla
}V\right)  ^{i}=-\left(  \widetilde{\nabla}W\right)  ^{i}\label{tilde}%
\end{equation}
with the affine parameter $d\widetilde{t}=dt/\widetilde{\sigma}$ and with
$\widetilde{\mathbf{J}}=\mathbf{J}_{1}\mathbf{\,J}_{2}^{-1}=\overline
{\mathbf{J}}^{-1}$ being a $J_{\widetilde{G}}$-tensor for the new metric
$\widetilde{G}=\widetilde{\sigma}\mathbf{J}_{2}G$ with $\widetilde{\sigma
}=\det(\mathbf{J}_{2})=\sigma_{2}$.

\begin{proposition}
The bi-cofactor system (\ref{bicoft}) has on $T^{\ast}Q$ the following
bi-quasihamiltonian representation:%
\begin{equation}
\frac{d}{dt}\left(
\begin{array}
[c]{c}%
q\\
p
\end{array}
\right)  =\frac{1}{\sigma_{1}}\Pi_{nc}(\mathbf{J}_{1})dH=\frac{1}{\sigma_{2}%
}\Pi_{nc}(\mathbf{J}_{2})dF,\label{bqh}%
\end{equation}
with two compatible noncanonical Poisson operators $\Pi_{nc}(\mathbf{J}_{1}) $
and $\Pi_{nc}(\mathbf{J}_{2})$ given by
\[
\Pi_{nc}(\mathbf{J})=\left(
\begin{array}
[c]{cc}%
0 & \mathbf{J}\\
-\mathbf{J}^{T} & \Omega
\end{array}
\right)  ,\ \ \ \Omega_{j}^{i}=\left(  \frac{\partial J_{i}^{k}}{\partial
q^{j}}-\frac{\partial J_{j}^{k}}{\partial q^{i}}\right)  p_{k}%
\]
and with the Hamiltonians%
\[
H=\frac{1}{2}p^{T}(\operatorname*{cof}\text{\thinspace}\mathbf{J}%
_{1})Gp+V(q),\text{ \ }F=\frac{1}{2}p^{T}(\operatorname*{cof}\text{\thinspace
}\mathbf{J}_{2})Gp+W(q).
\]

\end{proposition}

The representation (\ref{bqh}) follows directly from Proposition \ref{HAM}
applied independently to both cofactor representations of (\ref{bicoft}). The
fact that the operators $\Pi_{nc}(\mathbf{J}_{1})$ and $\Pi_{nc}%
(\mathbf{J}_{2})$ are compatible (i.e. that any linear combination $\eta
_{1}\Pi_{nc}(\mathbf{J}_{1})+\eta_{2}\Pi_{nc}(\mathbf{J}_{2})$ is Poisson) is
shown below. In the particular case of potential-cofactor systems (\ref{bar})
and (\ref{tilde}) this proposition yields their well-known
quasi-bi-Hamiltonian representation \cite{bro}, \cite{mt}.

\begin{theorem}
\label{CALKI}

\begin{enumerate}
\item The system (\ref{bqh}) has $n$ constants of motion%
\begin{equation}
H_{r}=E_{r}+V_{r}(q)=\frac{1}{2}p^{T}K_{r}Gp+V_{r}%
(q),\ \ \ \ \ \ \ \ r=1,...,n,\label{calki}%
\end{equation}
(with $H=H_{1}$ and $F=H_{n}$) where $K_{r}$ are $(1,1)$-Killing tensors (for
the metric $G$) defined by
\begin{equation}
\text{cof}(\mathbf{J}_{2}\mathbf{+}\xi\mathbf{J}_{1})=\sum_{i=0}^{n-1}%
K_{n-i}\xi^{i}\label{Ki}%
\end{equation}
$\mathbf{\ }$(so that $K_{1}=\operatorname*{cof}$\thinspace$\mathbf{J}%
_{1}\mathbf{,\ \ }K_{n}=\operatorname*{cof}$\thinspace$\mathbf{J}_{2}$) and
where the potentials $V_{r}$ can be obtained from two equivalent formulas%
\begin{equation}
\nabla V_{r}=\frac{1}{\sigma_{1}}K_{r}\mathbf{J}_{1}\nabla V_{1}\text{ \ \ or
\ }\nabla V_{r}=\frac{1}{\sigma_{2}}K_{r}\mathbf{J}_{2}\nabla V_{n},\text{
\ \ \ }V=V_{1},W=V_{n}.\label{Vrek}%
\end{equation}

\item The constants $H_{r}$ are in involution with respect to both operators
$\Pi_{nc}(\mathbf{J}_{1})$ and $\Pi_{nc}(\mathbf{J}_{2})$:%
\[
\left\{  H_{r},H_{s}\right\}  _{\Pi_{nc}(\mathbf{J}_{1})}=\left\{  H_{r}%
,H_{s}\right\}  _{\Pi_{nc}(\mathbf{J}_{2})}=0\text{ \ \ for all \ }%
r,s=1,\ldots,n.
\]

\end{enumerate}
\end{theorem}

To prove this theorem, we will first need

\begin{proposition}
\label{POMOC} In the variables $(\overline{q},\overline{p},\overline{t})$
related with $(q,p,t)$ through the map
\begin{equation}
\overline{q}=q\text{, \ }\overline{p}=\left(  \mathbf{J}_{1}^{T}\right)
^{-1}p\text{, \ }d\overline{t}=\frac{dt}{\sigma_{1}}\label{mapabar}%
\end{equation}
the system (\ref{bqh}) attains the quasi-bihamiltonian form%
\begin{equation}
\frac{d}{d\overline{t}}\left(
\begin{array}
[c]{c}%
\overline{q}\\
\overline{p}%
\end{array}
\right)  =\overline{\Pi}_{c}d\overline{H}=\frac{1}{\det(\overline{\mathbf{J}%
})}\overline{\Pi}_{nc}(\overline{\mathbf{J}})d\overline{F},\label{bqhbar}%
\end{equation}
with $\overline{H}=H$ and $\overline{F}=F$ (as functions on $T^{\ast}Q$) and
with
\begin{equation}
\overline{\Pi}_{c}=\Pi_{nc}(\mathbf{J}_{1}),\text{ \ \ }\overline{\Pi}%
_{nc}(\overline{\mathbf{J}})=\Pi_{nc}(\mathbf{J}_{2})\label{tosamo1}%
\end{equation}
(as tensors on $T^{\ast}Q$). Moreover, the tensor $\overline{\Pi}_{c}$ is
canonical in $(\overline{q},\overline{p})$-variables. Similarly, in the
variables $(\widetilde{q},\widetilde{p},\widetilde{t})$ defined by%
\begin{equation}
\widetilde{q}=q\text{, \ }\widetilde{p}=\left(  \mathbf{J}_{2}^{T}\right)
^{-1}p\text{, \ }d\widetilde{t}=\frac{dt}{\sigma_{2}}\text{.}\label{mapatilde}%
\end{equation}
(\ref{bqh}) attains the form%
\begin{equation}
\frac{d}{d\widetilde{t}}\left(
\begin{array}
[c]{c}%
\widetilde{q}\\
\widetilde{p}%
\end{array}
\right)  =\frac{1}{\det(\widetilde{\mathbf{J}})}\widetilde{\Pi}_{nc}%
(\widetilde{\mathbf{J}})d\widetilde{H}=\widetilde{\Pi}_{c}d\widetilde
{F},\label{bqhtilde}%
\end{equation}
with $\widetilde{H}=H$ and $\widetilde{F}=F$ (as functions on $T^{\ast}Q$) and
with
\begin{equation}
\widetilde{\Pi}_{c}=\Pi_{nc}(\mathbf{J}_{2}),\text{ \ \ }\widetilde{\Pi}%
_{nc}(\widetilde{\mathbf{J}})=\Pi_{nc}(\mathbf{J}_{1})\label{tosamo2}%
\end{equation}
(again considered as tensors on $T^{\ast}Q$) so that
\[
\widetilde{\Pi}_{c}=\overline{\Pi}_{nc}(\overline{\mathbf{J}}),\ \overline
{\Pi}_{c}=\widetilde{\Pi}_{nc}(\widetilde{\mathbf{J}}).
\]
Again, the tensor $\widetilde{\Pi}_{c}$ is canonical in $(\widetilde
{q},\widetilde{p})$-variables.
\end{proposition}

This proposition can be proved either by direct calculation or by observing
that the underlying bi-cofactor system (\ref{bicoft}) has in the variables
$(\overline{q},\overline{t})$ the potential-cofactor form (\ref{bar}) and in
the variables $(\widetilde{q},\widetilde{t})$ the cofactor-potential form
(\ref{tilde}) and using arguments similar to those used in the proof of
Proposition \ref{HAM}.

\begin{proof}
(of Theorem \ref{CALKI}). By Proposition \ref{POMOC}, the system (\ref{bqh})
has in variables $(\overline{q},\overline{p},\overline{t})$ the form
(\ref{bqhbar}) so that it is a so called Benenti system and therefore (see
\cite{ib}) $\overline{\Pi}_{c}=\Pi_{nc}(\mathbf{J}_{1})$, $\overline{\Pi}%
_{nc}(\overline{\mathbf{J}})=\Pi_{nc}(\mathbf{J}_{2})$ are compatible and the
system has $n$ constants of motion of the form%
\begin{equation}
\overline{H}_{r}=\overline{E}_{r}+\overline{V}_{r}(\overline{q})=\frac{1}%
{2}\overline{p}^{T}\overline{K}_{r}\overline{G}\overline{p}+\overline{V}%
_{r}(\overline{q}),\ \ \ \ \ \ \ \ r=1,...,n,\label{calkibar}%
\end{equation}
with $\overline{G}=\sigma_{1}\mathbf{J}_{1}G$ and where the Killing tensors
$\overline{K}_{r}$ of the metric $\overline{G}$ are determined by the
expansion%
\begin{equation}
\operatorname*{cof}\left(  \overline{\mathbf{J}}\mathbf{+}\xi I\right)
=\sum_{i=0}^{n-1}\overline{K}_{n-i}\xi^{i}\label{Kibar}%
\end{equation}
(so that $\overline{K}_{1}=I\mathbf{,\ \ }K_{n}=\operatorname*{cof}$%
\thinspace$\overline{\mathbf{J}}$) while $\overline{V}_{r}$ are separable
potentials satisfying
\begin{equation}
\overline{K}_{r}\overline{\nabla}\,\overline{V}_{1}=\overline{\nabla
}\,\overline{V}_{r}.\label{Vrekbar}%
\end{equation}
\bigskip On the other hand,
\[
\operatorname*{cof}\left(  \overline{\mathbf{J}}\mathbf{+}\xi I\right)
=\operatorname*{cof}\left(  \mathbf{J}_{2}\mathbf{J}_{1}^{-1}\mathbf{+}\xi
I\right)  =\operatorname*{cof}(\mathbf{J}_{1})^{-1}\operatorname*{cof}\left(
\mathbf{J}_{2}\mathbf{+}\xi\mathbf{J}_{1}\right)  =K_{1}^{-1}\sum_{i=0}%
^{n-1}K_{n-i}\xi^{i},
\]
so that by comparing with (\ref{Kibar}) we obtain
\begin{equation}
\overline{K}_{i}=K_{1}^{-1}K_{i},\ \ \ \ i=1,\ldots,n\label{kil1}%
\end{equation}
and
\[
\overline{E}_{r}=\frac{1}{2}\overline{p}^{T}\overline{K}_{r}\overline
{G}\overline{p}=\frac{1}{2}p^{T}\mathbf{J}_{1}^{-1}K_{1}^{-1}K_{r}\sigma
_{1}\mathbf{J}_{1}G\left(  \mathbf{J}^{-1}\right)  ^{T}p=E_{r}.
\]
The last equality follows from $\mathbf{J}_{1}^{-1}K_{1}^{-1}=\sigma_{1}I$ and
$\mathbf{J}_{1}G\left(  \mathbf{J}^{-1}\right)  ^{T}=\mathbf{J}_{1}%
\mathbf{J}^{-1}G=G$, so that indeed $\overline{E}_{r}=E_{r}$ if we define
$K_{i}$ as in (\ref{Ki}). Thus, if we put $V_{r}=\overline{V}_{r}$ we obtain
that $\overline{H}_{r}=H_{r}$ (as functions on $T^{\ast}Q$). Now, substituting
(\ref{kil1}) into (\ref{Vrekbar}) we get
\[
K_{1}^{-1}K_{r}\sigma_{1}J_{1}\nabla V_{1}=\sigma_{1}J_{1}\nabla V_{r}\text{
\ \ \ or \ \ \ }K_{r}J_{1}\nabla V_{1}=K_{1}J_{1}\nabla V_{r}%
\]
which yields the first formula in (\ref{Vrek}). Naturally, the functions
$H_{r}$ Poisson-commute with respect to both Poisson tensors $\Pi
_{nc}(\mathbf{J}_{i})$ in (\ref{bqh}) since $\overline{H}_{r}=H_{r}$
Poisson-commute with respect to both Poisson tensors $\overline{\Pi}%
_{c},\overline{\Pi}_{nc}(\overline{\mathbf{J}})$ in (\ref{bqhbar}) and since
these tensors are just $\Pi_{nc}(\mathbf{J}_{i})$ written in the variables
$(\overline{q},\overline{p})$, according to (\ref{tosamo1}). That proves all
the statements in Theorem \ref{CALKI} except the second formula in
(\ref{Vrek}). Consider now the system (\ref{bqhtilde}). It is also a Benenti
system so it also has $n$ constants of motion of the form
\begin{equation}
\widetilde{H}_{r}=\widetilde{E}_{r}+\widetilde{V}_{r}(\widetilde{q})=\frac
{1}{2}\widetilde{p}^{T}\widetilde{K}_{r}\widetilde{G}\widetilde{p}%
+\widetilde{V}_{r}(\widetilde{q}),\ \ \ \ \ \ \ \ r=1,...,n,\label{Htilde}%
\end{equation}
(with $\widetilde{G}=\sigma_{2}\mathbf{J}_{2}G$ and with $\widetilde
{H}=\widetilde{H}_{n},$ $\widetilde{F}=\widetilde{H}_{1}$) where the Killing
tensors $\widetilde{K}_{r}$ of the metric $\widetilde{G}$ are determined by%
\[
\operatorname*{cof}\left(  \widetilde{\mathbf{J}}\mathbf{+}\xi I\right)
=\sum_{i=0}^{n-1}\widetilde{K}_{n-i}\xi^{i}%
\]
(so that $\widetilde{K}_{1}=I$, $\widetilde{K}_{n}=\operatorname*{cof}%
\,\widetilde{\mathbf{J}}$) while $\widetilde{V}_{r}$ are separable potentials
satisfying
\begin{equation}
\widetilde{K}_{r}\widetilde{\nabla}V_{1}=\widetilde{\nabla}\widetilde{V}%
_{r}.\label{Vrektilde}%
\end{equation}
Repeating the procedure above we obtain an equivalent proof of Theorem
\ref{CALKI}. However, this time we get
\begin{align*}
\sum_{i=0}^{n-1}\widetilde{K}_{n-i}\xi^{i}  & =\operatorname*{cof}\left(
\widetilde{\mathbf{J}}\mathbf{+}\xi I\right)  =\operatorname*{cof}%
(\mathbf{J}_{1}\mathbf{J}_{2}^{-1}+\xi I)=K_{n}^{-1}\operatorname*{cof}%
(\mathbf{J}_{1}+\xi\mathbf{J}_{2})=\\
& =K_{n}^{-1}\xi^{n-1}%
{\textstyle\sum\limits_{i=0}^{n-1}}
K_{n-i}\xi^{-i}=K_{n}^{-1}%
{\textstyle\sum\limits_{i=0}^{n-1}}
K_{n-i}\xi^{n-i-1},
\end{align*}
which gives
\begin{equation}
\widetilde{K}_{i}=K_{n}^{-1}K_{n-i+1},\ \ \ \ \ i=1,\ldots,n.\label{kil2}%
\end{equation}
This also yields $\widetilde{E}_{i}=E_{n-i+1}$ and thus \ $\widetilde{V}%
_{i}=V_{n-i+1}$, $\widetilde{H}_{i}=H_{n-i+1}=\overline{H}_{n-i+1}$ for all
$i=1,\ldots,n$. By transforming the formula (\ref{Vrektilde}) to
$(q,p)$-coordinates (similarly as we did for bar-coordinates) we obtain the
second formula in (\ref{Vrek}). Finally, the functions $K_{i}$ defined by
(\ref{Ki}) must be $(1,1)$-Killing tensors for $G$ since cof$(\mathbf{J}%
_{2}\mathbf{+}\xi\mathbf{J}_{1})$ is a $(1,1)$-Killing tensor for any value of
the parameter $\xi$ and since Killing tensors of $G$ constitute a vector space.
\end{proof}

Note also that the direct map between variables $(\overline{q},\overline
{p},\overline{t})$ and $(\widetilde{q},\widetilde{p},\widetilde{t})$ is
obtained by composing the map (\ref{mapabar}) with the map (\ref{mapatilde}).
It attains the form%
\begin{equation}
\widetilde{q}=\overline{q},\text{ \ }\widetilde{p}=\left(  \overline
{\mathbf{J}}^{T}\right)  ^{-1}\overline{p},\text{ \ }d\widetilde{t}%
=\frac{d\overline{t}}{\det(\overline{\mathbf{J}})}\text{ \ or \ }\overline
{q}=\widetilde{q},\text{ \ }\overline{p}=\left(  \widetilde{\mathbf{J}}%
^{T}\right)  ^{-1}\widetilde{p},\text{ \ }d\overline{t}=\frac{d\widetilde{t}%
}{\det(\widetilde{\mathbf{J}})}.\label{bt}%
\end{equation}
Further, by comparing (\ref{kil1}) and (\ref{kil2}) we obtain
\[
\widetilde{K}_{i}=\overline{K}_{n}^{-1}\overline{K}_{n-i+1}.
\]
In the remaining part of this chapter will shortly discuss how the two
equivalent systems (\ref{bqhbar}) and (\ref{bqhtilde}) can be embedded in
quasi-bihamiltonian chains and discuss the relation between these chains.

Since the system (\ref{bqhbar}) has $n$ commuting with respect to both
operators $\overline{\Pi}_{c}$ and $\overline{\Pi}_{nc}$ integrals of motion
$\overline{H}_{r}$ it belongs to the set of $n$ commuting Hamiltonian vector
fields%
\begin{equation}
\frac{d}{d\overline{t}_{r}}\left(
\begin{array}
[c]{c}%
\overline{q}\\
\overline{p}%
\end{array}
\right)  =\overline{\Pi}_{c}\,d\overline{H}_{r}\equiv\overline{X}%
_{r},\ \ \ r=1,...,n,\label{hambar}%
\end{equation}
(where $\ d\overline{t}_{1}=d\overline{t}=dt/\det(\mathbf{J}_{1})$) and the
system (\ref{bqhbar}) itself defines the first vector field $\overline{X}_{1}
$. Similarly, since the system (\ref{bqhtilde}) has $n$ commuting with respect
to both $\widetilde{\Pi}_{c}$ and $\widetilde{\Pi}_{nc}$ integrals of motion
$\widetilde{H}_{r}$ it belongs to the set of $n$ commuting Hamiltonian vector
fields%
\begin{equation}
\frac{d}{d\widetilde{t}_{r}}\left(
\begin{array}
[c]{c}%
\widetilde{q}\\
\widetilde{p}%
\end{array}
\right)  =\widetilde{\Pi}_{c}\,d\widetilde{H}_{r}\equiv\widetilde{X}%
_{r},\ \ \ r=1,...,n,\label{hamtilde}%
\end{equation}
(where$\ d\widetilde{t}_{1}=d\widetilde{t}=dt/\det(\mathbf{J}_{2})$ so that
$d\widetilde{t}_{1}=d\overline{t}_{1}/\det(\overline{\mathbf{J}})$) and it
also is the first vector field $\widetilde{X}_{1}$. By the above construction,
the vector fields $\overline{X}_{1}$ and $\widetilde{X}_{1}$ are parallel
\[
\widetilde{X}_{1}=\det(\overline{\mathbf{J}})\,\overline{X}_{1}\text{ \ \ \ or
\ \ }\overline{X}_{1}=\det(\widetilde{\mathbf{J}})\,\widetilde{X}_{1}%
\]
which once again reflects the geodesic equivalence of the systems (\ref{bar})
and (\ref{tilde}) on $Q$.

Moreover, vector fields (\ref{hambar}) belong to the following
quasi-bihamiltonian chain:%
\begin{align*}
\overline{X}_{1}  & =\overline{\Pi}_{c}\,d\overline{H}_{1}=\frac{1}%
{\overline{\rho}_{n}}\overline{\Pi}_{nc}(\overline{\mathbf{J}})d\overline
{H}_{n}\\
\overline{X}_{r}  & =\overline{\Pi}_{c}\,d\overline{H}_{r}=\frac
{\overline{\rho}_{r-1}}{\overline{\rho}_{n}}\overline{\Pi}_{nc}(\overline
{\mathbf{J}})d\overline{H}_{n}-\overline{\Pi}_{nc}(\overline{\mathbf{J}%
})d\overline{H}_{r-1},\ \ \ r=2,...,n,
\end{align*}
where the functions $\overline{\rho}_{r}$ are defined through the polynomial
expansion of $\det(\overline{\mathbf{J}}+\xi I)$:%
\[
\ \ \ \ \ \ \det(\overline{\mathbf{J}}+\xi I)=\sum_{i=0}^{n}\overline{\rho
}_{i}\xi^{n-i}\text{ \ \ }%
\]
(so that $\overline{\rho}_{n}=\det\overline{\mathbf{J}}$). Similarly, vector
fields (\ref{hamtilde}) belong to a similar quasi-bihamiltonian chain:%
\begin{align*}
\widetilde{X}_{1}  & =\widetilde{\Pi}_{c}\,d\widetilde{H}_{1}=\frac
{1}{\widetilde{\rho}_{n}}\widetilde{\Pi}_{nc}(\widetilde{\mathbf{J}%
})d\overline{H}_{n}\\
\widetilde{X}_{r}  & =\widetilde{\Pi}_{c}\,d\widetilde{H}_{r}=\frac
{\widetilde{\rho}_{r-1}}{\widetilde{\rho}_{n}}\widetilde{\Pi}_{nc}%
(\widetilde{\mathbf{J}})d\widetilde{H}_{n}-\widetilde{\Pi}_{nc}(\widetilde
{\mathbf{J}})d\widetilde{H}_{r-1},\ \ \ r=2,...,n,
\end{align*}
where $\widetilde{\rho}_{r}$ are defined through%
\[
\det(\widetilde{\mathbf{J}}+\xi I)=\sum_{i=0}^{n}\widetilde{\rho}_{i}\xi^{n-i}%
\]
(so that $\widetilde{\rho}_{n}=\det\widetilde{\mathbf{J}}$). Since
$\widetilde{\mathbf{J}}=\overline{\mathbf{J}}^{-1}$ we have that
$\overline{\rho}_{r}$ and $\widetilde{\rho}_{r}$ are related via
\[
\widetilde{\rho}_{r}=\frac{\overline{\rho}_{n-r}}{\overline{\rho}_{n}}\text{
\ \ or \ }\overline{\rho}_{r}=\frac{\widetilde{\rho}_{n-r}}{\widetilde{\rho
}_{n}}.
\]
Comparing both chains we obtain that the vector fields $\overline{X}_{r}$ and
$\widetilde{X}_{r}$ are related through%
\[
\widetilde{X}_{1}=\overline{\rho}_{n}\,\overline{X}_{1},\text{ \ \ \ \ }%
\widetilde{X}_{i}=\overline{\rho}_{n-i+1}\overline{X}_{1}-\overline{X}%
_{n-i+2},\text{ \ \ }i=2,\ldots,n.\text{\ }%
\]

\section{Flat bi-cofactor systems.}

Let us recall that a pseudoriemannian space is called space of constant
curvature if the curvature tensor $R_{ijkl}$ has the form%
\begin{equation}
R_{ijkl}=K\left(  g_{jl}g_{ik}-g_{jk}g_{il}\right) \label{Riemann}%
\end{equation}
for some scalar function $K\,$. By Bianchi identity it follows then that $K$
is a constant, related to scalar (Ricci) curvature $\varkappa=R_{ik}g^{ik}$
through $\varkappa=Kn(n-1)$. Thus, for such spaces the condition $\varkappa=0$
or $K=0$ implies that the Riemann tensor $R_{ijkl}$ is zero i.e. that the
metric $g$ is flat.

Suppose now that $g$ is a metric of constant curvature and that $\overline{g}
$ is another metric tensor obtained by deforming $g$ through%
\begin{equation}
\overline{G}=\sigma\mathbf{J}G\label{deformg}%
\end{equation}
(with $\mathbf{J}$ being a $J_{G}$-tensor $\mathbf{J}$ and with $\sigma
=\det\mathbf{J}$). Then, by the classical result of Beltrami
\textbf{\cite{Beltrami} }we know that $\overline{g}$ is also of constant
curvature. Moreover, for two metrics $g$ and $\overline{g}$ that are
geodesically equivalent and of constant curvature their scalar curvatures
$\varkappa$ and $\overline{\varkappa}$ are related by the formula
\begin{equation}
\overline{\varkappa}\,\overline{g}_{ij}=\varkappa g_{ij}-\nabla_{i}f_{j}%
+f_{i}f_{j}\label{zwkappa}%
\end{equation}
(see \cite{Schouten} p. 293) where the covector $f_{i}$ is defined as
\begin{equation}
f_{i}=\frac{1}{2(n+1)}\frac{\partial}{\partial q_{i}}\left(  \ln\frac
{\det\overline{g}}{\det g}\right)  .\label{f}%
\end{equation}
A simple calculation shows that for our choice of $g,\overline{g}$ we have%
\[
f_{i}=-\frac{1}{n+1}\sigma_{i}\text{ where }\sigma_{i}=\frac{1}{\sigma}%
\frac{\partial\sigma}{\partial q_{i}}.
\]
Substituting this into (\ref{zwkappa}) and performing contraction with
$\overline{G}$ we obtain%
\[
\overline{\varkappa}=\frac{\sigma}{n}\left[  \varkappa\operatorname*{tr}%
\mathbf{J}+\tfrac{1}{n+1}J^{ij}\left(  \sigma_{i}\sigma_{j}+\tfrac{1}%
{n+1}\nabla_{i}\sigma_{j}\right)  \right]  .
\]
(the summation convention applies as usual). Thus, we see that if
$\varkappa=0$ then a sufficient condition for $\overline{\varkappa}$ to be
zero is
\begin{equation}
J^{ij}\left(  \sigma_{i}\sigma_{j}+\tfrac{1}{n+1}\nabla_{i}\sigma_{j}\right)
=0.\label{wk}%
\end{equation}
Let us now assume that the metric $G$ of the system (\ref{bicoft}) is flat
(i.e. $\varkappa=0$) so that in some coordinate system $(q^{i})$ it assumes
the form%
\begin{equation}
G=\operatorname*{diag}(\varepsilon_{1},\ldots,\varepsilon_{n})\text{ with
}\varepsilon_{i}=\pm1\text{.}\label{Gflat}%
\end{equation}
(note that then $g=G^{-1}=\operatorname*{diag}(\varepsilon_{1},\ldots
,\varepsilon_{n})$ in this particular coordinate system too while $\Gamma
_{jk}^{i}=0$). Suppose now that we want to "deform" this system as in
Proposition \ref{czast3} by introducing the new independent variable
$dt_{3}=dt/\sigma$ where $\sigma=\det\mathbf{J}$ for some new $J_{G}$-tensor
$\mathbf{J}$ but in such a way that the resulting equivalent metric
$\overline{G}=\sigma\mathbf{J}G$ is also flat so that the geodesically
equivalent system (\ref{bicoft3}) is a flat Newton system (in this section we
will use $\mathbf{J}$, $\sigma$ and $\overline{G}$ instead of $\mathbf{J}%
_{3},$ $\det\mathbf{J}_{3}$ and $G_{3}$ to shorten the notation). A sufficient
condition for doing this is to take $\mathbf{J}$ that satisfies (\ref{wk}). In
the Cartesian (with respect to $g$) coordinates $(q^{i})$ the contravariant
form of the tensor $\mathbf{J}$ is given by (\ref{og}). However, by Theorem
B.4.3 in \cite{be1} we know that $m=0$ or else $\overline{\varkappa}\neq0$.
Thus, our aim is to find a more explicit form of the condition (\ref{wk}) for
$\mathbf{J}$ given by (\ref{og}) with $m=0$.

Let us for the moment denote the $(2,0)$-form of $\mathbf{J}$ as given in
(\ref{og}) by $\mathbf{J}_{c}$ ($\mathbf{J}$-contravariant) so that
$\mathbf{J}_{c}=\mathbf{J}G$ or $\mathbf{J}=\mathbf{J}_{c}g$. We have then

\begin{theorem}
Assume that $G$ is of the form (\ref{Gflat}) and that $\mathbf{J}$ is a
$J_{G}$-tensor such that its contravariant form $\mathbf{J}_{c}$ is given by
(\ref{og}). Then for the metric $\overline{G}=\sigma\mathbf{J}G$ to be flat it
is sufficient that $m=0$ and%
\begin{equation}
\beta^{T}g\left(  \operatorname*{cof}\mathbf{J}\right)  \beta=0\text{ or
}\beta^{T}\left(  \operatorname*{cof}\mathbf{J}_{c}\right)  \beta
=0.\label{wkmac}%
\end{equation}

\end{theorem}

\begin{proof}
Both conditions in (\ref{wkmac}) are equivalent since $\operatorname*{cof}%
\mathbf{J}_{c}=\operatorname*{cof}(\mathbf{J}G)=\operatorname*{cof}%
G\,\operatorname*{cof}\mathbf{J}=\det(G)\,g\,\operatorname*{cof}\mathbf{J}$.
We have to show that the condition (\ref{wk}) in our setting attains the form
(\ref{wkmac}). Since $\mathbf{J}$ is torsionless it satisfies the identity
\[
\sigma\frac{\partial\left(  \operatorname*{tr}\mathbf{J}\right)  }{\partial
q^{i}}\mathbf{=J}_{i}^{h}\frac{\partial\sigma}{\partial q^{h}},
\]
or in the matrix form
\begin{equation}
\sigma\mathbf{\,}d(\operatorname*{tr}\mathbf{J})=\mathbf{J}^{T}d\sigma
.\label{tozs}%
\end{equation}
Since $J_{j}^{i}=J^{ik}g_{kj}=\beta^{i}\varepsilon_{j}q^{j}+\beta
^{j}\varepsilon_{j}q^{i}+\gamma^{ij}\varepsilon_{j}$ (no summation) we have
$\operatorname*{tr}(\mathbf{J})=J_{i}^{i}=2\beta^{i}\varepsilon_{i}%
q^{i}+\gamma^{ii}\varepsilon_{i}$ so that $d(\operatorname*{tr}\mathbf{J}%
)=(\varepsilon_{1}\beta^{1},\ldots,\varepsilon_{n}\beta^{n})^{T}$. Thus,
(\ref{tozs}) reads%
\begin{equation}
d\sigma=2\left(  \operatorname*{cof}\mathbf{J}\right)  ^{T}g\beta=2g\left(
\operatorname*{cof}\mathbf{J}\right)  \beta.\label{grads}%
\end{equation}
Therefore%
\begin{equation}
\sigma_{i}J^{ij}\sigma_{j}=\frac{1}{\sigma^{2}}\frac{\partial\sigma}{\partial
q^{i}}J^{ij}\frac{\partial\sigma}{\partial q^{j}}=\frac{1}{\sigma^{2}}%
(d\sigma)^{T}\,\mathbf{J}\,d\sigma=\frac{4}{\sigma^{2}}\beta^{T}g\left(
\operatorname*{cof}\mathbf{J}\right)  \beta.\label{pierwszy}%
\end{equation}
Further%
\[
J^{ij}\nabla_{i}\sigma_{j}=J^{ij}\frac{\partial}{\partial q^{i}}\left(
\frac{1}{\sigma}\frac{\partial\sigma}{\partial q^{j}}\right)  =-\sigma
_{i}J^{ij}\sigma_{j}+\frac{1}{\sigma}J^{ij}\frac{\partial^{2}\sigma}{\partial
q^{i}\partial q^{j}}.
\]
But, using (\ref{grads}) twice and (\ref{Gflat}) we obtain%
\begin{align*}
J^{ij}\frac{\partial^{2}\sigma}{\partial q^{i}\partial q^{j}}  & =J^{ij}%
\frac{\partial}{\partial q^{i}}\left(  \frac{\partial\sigma}{\partial q^{j}%
}\right)  =2J^{ij}g_{jk}\frac{\partial}{\partial q^{i}}\left(
\operatorname*{cof}\mathbf{J}\right)  _{s}^{k}\,\beta^{s}=2J_{k}^{i}%
\frac{\partial}{\partial q^{i}}\left(  \operatorname*{cof}\mathbf{J}\right)
_{s}^{k}\,\beta^{s}=\\
& =2\left[  \frac{\partial}{\partial q^{i}}\left(  J_{k}^{i}\left(
\operatorname*{cof}\mathbf{J}\right)  _{s}^{k}\right)  -\,\left(
\operatorname*{cof}\mathbf{J}\right)  _{s}^{k}\frac{\partial}{\partial q^{i}%
}J_{k}^{i}\right]  \beta^{s}=2\frac{\partial\sigma}{\partial q^{i}}\beta
^{i}-2(n+1)\beta^{k}\varepsilon_{k}\left(  \operatorname*{cof}\mathbf{J}%
\right)  _{s}^{k}\beta^{s}=\\
& =4\beta^{T}g\left(  \operatorname*{cof}\mathbf{J}\right)  \beta
-2(n+1)\beta^{T}g\left(  \operatorname*{cof}\mathbf{J}\right)  \beta.
\end{align*}
so that%
\[
J^{ij}\frac{\partial^{2}\sigma}{\partial q^{i}\partial q^{j}}=2(1-n)\beta
^{T}\,g(\operatorname*{cof}\mathbf{J})\,\beta.
\]
Thus,%
\begin{equation}
J^{ij}\nabla_{i}\sigma_{j}=-\frac{2}{\sigma}(1+n)\beta^{T}%
\,g(\operatorname*{cof}\mathbf{J})\,\beta.\label{drugi}%
\end{equation}
Plugging (\ref{pierwszy}) and (\ref{drugi}) into (\ref{wk}) we immediately
obtain (\ref{wkmac}).
\end{proof}

Therefore, we have showed that for any flat bi-cofactor system (\ref{bicoft})
there exists a multi-parameter family (with $\frac{1}{2}n(n+3)-1$ parameters)
of geodesically equivalent (but algebraically very different) flat bi-cofactor systems.

\begin{remark}
The condition (\ref{wkmac}) can be written as%
\begin{equation}
\beta^{T}g\,\left(  \operatorname*{cof}\gamma g\right)  \mathbf{\,}%
\beta=0\text{ or }\beta^{T}\,\left(  \operatorname*{cof}\gamma\right)
\mathbf{\,}\beta=0.\label{wksimple}%
\end{equation}

\end{remark}

\section{Separation curves for bi-cofactor systems}

A system of $n$ algebraic equations of the form
\begin{equation}
\varphi_{i}(\lambda^{i},\mu_{i};a_{1},...,a_{n}%
)=0,\ \ \ \ \ i=1,...,n,\ \ \ \det\left[  \frac{\partial\varphi_{i}}{\partial
a_{j}}\right]  \neq0,\label{sr}%
\end{equation}
each containing only one pair $(\lambda^{i},\mu_{i})$ of coordinates
$(\lambda,\mu)$ on $T^{\ast}Q$ (and with real coefficients $a_{i}$) is called
\emph{separation relations}. The condition in (\ref{sr}) means that we can
solve the equations (\ref{sr}) with respect to $a_{i}$ obtaining $n$
independent functions on\ $T^{\ast}Q$ of the form $a_{i}=H_{i}(\lambda,\mu)$,
$i=1,\ldots,n$. If the functions $W_{i}(\lambda^{i},a)~$\ are solutions of a
system of $n$ decoupled \ ODE's
\begin{equation}
\varphi_{i}\left(  \lambda^{i},\mu_{i}=\frac{dW_{i}(\lambda^{i},a)}%
{d\lambda^{i}},a_{1},\ldots,a_{n}\right)  =0\text{, \ }i=1,...,n,\label{char}%
\end{equation}
then the function $W(\lambda,a)=%
{\textstyle\sum\nolimits_{i=1}^{n}}
W_{i}(\lambda^{i},a)$ is a solution of all the equations (\ref{char}) and
simultaneously it is an additively separable solution of all Hamilton-Jacobi equations%

\begin{equation}
H_{i}\left(  \lambda^{1},...,\lambda^{n},\frac{\partial W}{\partial\lambda
^{1}},...,\frac{\partial W}{\partial\lambda^{n}}\right)  =a_{i}%
,\ \ \ \ \ \ \ i=1,...,n.\label{HJ}%
\end{equation}
simply because solving (\ref{sr}) to the form $a_{i}=H_{i}(\lambda,\mu)$ is a
purely algebraic operation. The Hamiltonians $H_{i}$ Poisson-commute by the
classical theorem of Jacobi. The function $W(\lambda,a)$ is a generating
function for the canonical transformation $(\lambda,\mu)\rightarrow(b,a)$ to
the new set of coordinates that simultaneously linearize all the Hamiltonian
equations%
\begin{equation}
u_{t_{i}}=\Pi_{c}\,dH_{i}=X_{H_{i}},\ \ \ \ \ \ \ i=1,...,n.\label{hamflow}%
\end{equation}
The coordinates $(\lambda,\mu)$ are thus called the \emph{separation
coordinates} for the Hamiltonian systems (\ref{hamflow}).

In the case that the relations (\ref{sr}) are affine in $a_{i}$ the obtained
systems belong to the well-known class of (generalized) St\"{a}ckel separable systems.

Let us now consider a special subclass of St\"{a}ckel systems given by the
following separation relations:
\begin{equation}
H_{1}(\lambda^{i})^{n-1}-H_{2}(\lambda^{i})^{n-2}+...+(-1)^{n-1}H_{n}=\frac
{1}{2}f_{i}(\lambda^{i})\mu_{i}^{2}+\gamma_{i}(\lambda^{i}%
),\ \ \ \ i=1,...,n,\label{Benen}%
\end{equation}
where $f_{i}$ and $\gamma_{i}$ are smooth functions. Such systems are known as
Benenti systems. In the particular case that $f_{i}(\lambda^{i})=f(\lambda
^{i})$ and $\gamma_{i}(\lambda^{i})=\gamma(\lambda^{i})$, separation relations
(\ref{Benen}) are given by $n$ copies of the so called separation curve
\begin{equation}
H_{1}\lambda^{n-1}-H_{2}\lambda^{n-2}+...+(-1)^{n-1}H_{n}=\frac{1}{2}%
f(\lambda)\mu^{2}+\gamma(\lambda).\label{SK}%
\end{equation}
so that now $\lambda$ and $\mu\in\mathbf{R}$. By solving the system of $n$
copies of this relation (with $i$-th copy containing variables labelled
$(\lambda_{i},\mu_{i})$) with respect to $H_{i}$ we find that the Hamiltonians
$H_{i}$ attain the form%

\[
H_{r}=E_{r}+V_{r}(\lambda)=\frac{1}{2}\mu^{T}K_{r}G\mu+V_{r}(\lambda
),\ \ \ \ \ \ \ \ r=1,...,n,
\]
(cf (\ref{calki})) with the metric tensor
\[
G=\operatorname*{diag}\left(  \frac{f(\lambda^{1})}{\Delta_{1}},...,\frac
{f(\lambda^{n})}{\Delta_{n}}\right)
\]
where $\Delta_{i}=%
{\textstyle\prod\limits_{j\neq i}}
(\lambda^{i}-\lambda^{j})$ while the $(1,1)$-tensors $K_{r}$ are generated by
the expansion
\[
\operatorname*{cof}(\mathbf{J+}\xi I)=\sum_{i=0}^{n-1}K_{n-i}\xi^{i}%
\]
with the $J_{G}$-tensor $\mathbf{J}=\operatorname*{diag}(\lambda_{1}%
,\ldots,\lambda_{n})$. Thus \cite{bma}:%
\[
K_{r}=\operatorname*{diag}\left(  \frac{\partial\rho_{r}}{\partial\lambda^{1}%
},...,\frac{\partial\rho_{r}}{\partial\lambda^{n}}\right)
\]
where the functions $\rho_{r}$ can be obtained from
\[
\ \ \ \ \ \ \det(\mathbf{J}+\xi I)=\sum_{i=0}^{n}\rho_{i}\xi^{n-i}.\text{
\ \ }%
\]
Coordinate-free expression for $K_{r}$ is as follows \cite{ben}
\[
K_{r+1}=\rho_{r}I-\mathbf{J}K_{r},\text{ }r=0,1,\ldots,n-1\text{, \ }\rho
_{0}=1,\ \text{\ }K_{0}=0,
\]
or alternatively
\[
K_{r}=%
{\displaystyle\sum\limits_{k=0}^{r-1}}
\rho_{k}\,\left(  -\mathbf{J}\right)  ^{r-1-k}\text{, \ \ \ \ }r=1,\ldots,n.
\]
It is important to stress that in case that eigenvalues of $J$ are not simple
the obtained tensors $K_{r}$ will not be independent and thus will not
generate an integrable system (see also below).

For a particular choice $\gamma(\lambda^{i})=(\lambda^{i})^{k},\ k\in
\mathbf{Z}$ in the separation curve (\ref{SK}) we obtain a family of separable
potentials that can be constructed recursively by \cite{bma}%

\begin{equation}
V_{r}^{(k+1)}=\rho_{r}V_{1}^{(k)}-V_{r+1}^{(k)}\text{ with }V_{r}%
^{(0)}=(-1)^{n-1}\delta_{rn}\text{.}\label{a}%
\end{equation}
This recursion can be reversed%
\begin{equation}
V_{r}^{(k-1)}=\frac{\rho_{r-1}}{\rho_{n}}V_{n}^{(k)}-V_{r-1}^{(k)}%
\text{.}\label{b}%
\end{equation}
In both cases we put $V_{r}^{(k)}=0$ for $r<0$ or $r>n$. The above recursion
can be written in a matrix form as
\begin{equation}
V^{(k)}(\lambda)=R^{k}(\lambda)V^{(0)},\ \ \ \ k\in Z,\label{rekV}%
\end{equation}
where $V^{(k)}(\lambda)=(V_{1}^{(k)}(\lambda),...,V_{n}^{(k)}(\lambda
))^{T},\ V^{(0)}=(0,...,0,(-1)^{n-1})^{T}$ and
\begin{equation}
R=\left(
\begin{array}
[c]{ccccc}%
\rho_{1}(\lambda) & -1 & 0 & \cdots & 0\\
\rho_{2}(\lambda) & 0 & -1 & \cdots & 0\\
\vdots & \vdots & \vdots & \cdots & \vdots\\
\rho_{n-1}(\lambda) & 0 & 0 & \cdots & -1\\
\rho_{n}(\lambda) & 0 & 0 & \cdots & 0
\end{array}
\right)  .\label{R}%
\end{equation}
This recursion is equivalent to (\ref{a}) and (\ref{b}) and is invariant with
respect to any point change of variables on $Q$ as $R$ in (\ref{R}) is
expressed by coefficients of the characteristic polynomial of $\mathbf{J}$.
The first nontrivial potentials in the positive hierarchy are $V_{r}%
^{(n)}(\lambda)=\rho_{r}(\lambda),$ while for the negative hierarchy
$V_{r}^{(-1)}=\rho_{r-1}(\lambda)/\rho_{n}(\lambda).$

Let us now once again consider our systems (\ref{bar}) and (\ref{tilde}) and
their Hamiltonian formulations (\ref{hambar}) and (\ref{hamtilde})
respectively. From now on we will additionally assume that the tensor
$\overline{\mathbf{J}}$ (and hence $\widetilde{\mathbf{J}}$) has all its
eigenvalues real and \emph{simple} (i.e. both are so called $L$-tensors
\cite{be1}). Then the tensors $\overline{K}_{r}$ (and $\widetilde{K}_{r}$
likewise) are independent and thus (\ref{hambar}) and (\ref{hamtilde}) are
integrable. Moreover, both systems belong to the class of separable (in the
sense of Hamilton-Jacobi theory) systems called Benenti systems. It is known
that all Hamiltonian flows in (\ref{hambar}) are separable in variables
$(\overline{\lambda},\overline{\mu})$ where the new coordinates $\overline
{\lambda}^{i}$ are obtained from the characteristic equation of $\overline
{\mathbf{J}}$:
\begin{equation}
\det(\overline{\mathbf{J}}+\overline{\lambda}I)=0\label{lb}%
\end{equation}
(i.e. are (signed) eigenvalues of $\overline{\mathbf{J}}$) while the
corresponding momenta $\overline{\mu}^{i}$ are obtained from $\overline{\mu}=$
$(\Phi^{\prime-1})^{T}\overline{p}$ where $\Phi^{\prime}$ is the Jacobi matrix
of the map $\Phi:q\rightarrow\overline{\lambda}$ given by (\ref{lb}).
Similarly, all the flows in (\ref{hamtilde}) are separable in variables
$(\widetilde{\lambda},\widetilde{\mu})$ where $\widetilde{\lambda}^{i}$ are
obtained from
\begin{equation}
\det(\widetilde{\mathbf{J}}+\widetilde{\lambda}I)=0\label{lt}%
\end{equation}
with the corresponding momenta $\widetilde{\mu}^{i}$ obtained by
$\widetilde{\mu}=$ $(\Psi^{\prime-1})^{T}\widetilde{p}$ where $\Psi^{\prime}$
is the Jacobi matrix of the map $\Psi:q\rightarrow\widetilde{\lambda}$ given
by (\ref{lt}).

\begin{theorem}
The separation variables $(\overline{\lambda},\overline{\mu})$ of
(\ref{hambar}) and the separation variables $(\widetilde{\lambda}%
,\widetilde{\mu})$ of (\ref{hamtilde}) are related by the transformation%
\begin{equation}
\widetilde{\lambda}^{i}=\frac{1}{\overline{\lambda}^{i}},\ \ \ \widetilde{\mu
}_{i}=-\overline{\lambda}^{i}\overline{\mu}_{i}\text{ \ (no summation),\ \ \ }%
i=1,\ldots,n.\label{maplm}%
\end{equation}

\end{theorem}

\begin{proof}
By comparing (\ref{lb}) with (\ref{lt}) we obtain that $\overline{\lambda}%
^{i}=\frac{1}{\widetilde{\lambda}^{i}}$. The map between momenta
$\widetilde{\mu}$ and $\overline{\mu}$ can be found in the following way. We
know (cf. (\ref{mapabar}) and (\ref{mapatilde})) that $\overline{p}=\left(
\mathbf{J}_{1}^{T}\right)  ^{-1}p$ and $\widetilde{p}=\left(  \mathbf{J}%
_{2}^{T}\right)  ^{-1}p$ which yields $\widetilde{p}=\left(  \overline
{\mathbf{J}}^{T}\right)  ^{-1}\overline{p}$. Thus%
\[
\widetilde{\mu}=(\Psi^{\prime-1})^{T}\widetilde{p}=(\Psi^{\prime-1}%
)^{T}(\overline{\mathbf{J}}^{T})^{-1}\Phi^{\prime}{}^{T}\overline{\mu}%
=\Phi^{\prime}\overline{\mathbf{J}}^{-1}(\Psi^{\prime-1})^{T}\overline{\mu}.
\]
Since $\overline{\lambda}^{i}=1/\widetilde{\lambda}^{i}$ we see that
$\Psi^{\prime}=\Theta\Phi^{\prime}$ where
\[
\Theta=-\operatorname*{diag}\left(  \frac{1}{(\overline{\lambda}^{1})^{2}%
},\ldots,\frac{1}{(\overline{\lambda}^{n})^{2}}\right)
\]
so that $\Psi^{\prime-1}=-\Phi^{\prime-1}\operatorname*{diag}\left(
(\overline{\lambda}^{1})^{2},\ldots,(\overline{\lambda}^{n})^{2}\right)  $.
Inserting it in the above formula yields%
\[
\widetilde{\mu}=-\left(  \Phi^{\prime}\overline{\mathbf{J}}^{-1}\Phi
^{\prime-1}\operatorname*{diag}((\overline{\lambda}^{1})^{2},\ldots
,(\overline{\lambda}^{n})^{2})\right)  ^{T}\overline{\mu}.
\]
But $\Phi^{\prime}\overline{\mathbf{J}}^{-1}\Phi^{\prime-1}%
=\operatorname*{diag}(1/\overline{\lambda}^{1},\ldots,1/\overline{\lambda}%
^{n})$ since it is the inverse of the $L$-tensor $\overline{\mathbf{J}}$
written in its separation coordinates $\overline{\lambda}$. Inserting it into
the above formula we get the map between momenta as in (\ref{maplm}).
\end{proof}

According to the remarks above, the Benenti system (\ref{bqhbar}) in variables
$(\overline{\lambda},\overline{\mu})$ has the separation curve%
\begin{equation}
\overline{H}_{1}\overline{\lambda}^{n-1}-\overline{H}_{2}\overline{\lambda
}^{n-2}+...+(-1)^{n-1}\overline{H}_{n}=\frac{1}{2}\overline{f}(\overline
{\lambda})\overline{\mu}^{2}+\overline{\gamma}(\overline{\lambda
}).\label{SKbar}%
\end{equation}
Similarly, the separation curve for the Benenti system (\ref{bqhtilde}) is%
\begin{equation}
\widetilde{H}_{1}\widetilde{\lambda}^{n-1}-\widetilde{H}_{2}\widetilde
{\lambda}^{n-2}+...+(-1)^{n-1}\widetilde{H}_{n}=\frac{1}{2}\widetilde
{f}(\widetilde{\lambda})\widetilde{\mu}^{2}+\widetilde{\gamma}(\widetilde
{\lambda}).\label{SKtilde}%
\end{equation}
Applying the map (\ref{maplm}) to the separation curve (\ref{SKtilde}), using
that $\overline{H}_{r}=\widetilde{H}_{n-r+1}$ and comparing the result with
(\ref{SKbar}) we obtain
\begin{align*}
\overline{H}_{1}\overline{\lambda}^{n-1}-\overline{H}_{2}\overline{\lambda
}^{n-2}+...+(-1)^{n-1}\overline{H}_{n}  & =\frac{(-1)^{n-1}}{2}\widetilde
{f}\left(  \overline{\lambda}^{-1}\right)  \overline{\lambda}^{n+1}%
\overline{\mu}_{i}^{2}\\
& \\
& +(-1)^{n-1}\widetilde{\gamma}(\overline{\lambda}^{-1})\overline{\lambda
}^{n-1}.
\end{align*}

\begin{corollary}
If the functions $\overline{f},\widetilde{f}$ and $\overline{\gamma
},\widetilde{\gamma}$ satisfy the conditions:%
\begin{equation}
\overline{f}(\xi)=(-1)^{n-1}\widetilde{f}\left(  \xi^{-1}\right)  \xi
^{n+1},\text{ \ \ }\overline{\gamma}(\xi)=(-1)^{n-1}\widetilde{\gamma}%
(\xi^{-1})\xi^{n-1},\text{ \ \ }\xi\in\mathbf{R}\label{relf}%
\end{equation}
then the separation curves (\ref{SKbar}) and (\ref{SKtilde}) generate two
geodesically equivalent systems of Benenti type parametrized by two different
evolution parameters $\overline{t}$ and $\widetilde{t}$ such that
$d\widetilde{t}=d\overline{t}/\det(\overline{\sigma})$, where $\overline
{\sigma}=\rho_{n}=%
{\textstyle\prod\nolimits_{i=1}^{n}}
\overline{\lambda}^{i}$. The corresponding families of separable potentials
(\ref{rekV}) for both systems are related by
\[
\overline{V}^{(k)}=(-1)^{n-1}\,\widetilde{V}^{(n-k-1)}\text{ \ or
\ \ }\widetilde{V}^{(k)}=(-1)^{n-1}\,\overline{V}^{(n-k-1)}\text{ \ \ \ \ for
all }r\in\mathbf{Z}.
\]

\end{corollary}

It is known that the metric
\[
\overline{G}=\operatorname*{diag}\left(  \frac{\overline{f}(\overline{\lambda
}^{1})}{\overline{\Delta}_{1}},...,\frac{\overline{f}(\overline{\lambda}^{n}%
)}{\overline{\Delta}_{n}}\right)
\]
is of constant curvature if and only if $\overline{f}(\overline{\lambda}%
)=\sum_{k=0}^{n+1}c_{k}\overline{\lambda}^{k}\ $for some constants $c_{k}$.
From (\ref{relf}) it follows immediately that the equivalent metric
\[
\widetilde{G}=\operatorname*{diag}\left(  \frac{\widetilde{f}(\widetilde
{\lambda}^{1})}{\widetilde{\Delta}_{1}},...,\frac{\widetilde{f}(\widetilde
{\lambda}^{n})}{\widetilde{\Delta}_{n}}\right)
\]
is also of constant curvature, as in this case
\[
\widetilde{f}(\widetilde{\lambda})=(-1)^{n-1}\overline{f}(\widetilde{\lambda
}^{-1})\widetilde{\lambda}^{n+1}=(-1)^{n-1}\sum_{k=0}^{n+1}c_{k}%
\widetilde{\lambda}^{n-k+1}=(-1)^{n-1}\sum_{k=0}^{n+1}c_{n-k+1}\widetilde
{\lambda}^{k}.
\]

\section{Example: flat bi-cofactor systems geodesically equivalent to
Henon-Heiles system}

Let us illustrate the ideas of this paper on the example of the integrable
case of the Henon-Heiles system. It has the potential-cofactor form
(\ref{bar}):
\begin{equation}
\frac{d^{2}\overline{q}}{d\overline{t}^{2}}=-\overline{\nabla}V=-\left(
\operatorname*{cof}\overline{\mathbf{J}}\right)  ^{-1}\overline{\nabla
}W=-\left(
\begin{array}
[c]{c}%
3(\overline{q}^{1})^{2}+\frac{1}{2}\left(  \overline{q}^{2}\right)  ^{2}\\
\overline{q}^{1}\overline{q}^{2}%
\end{array}
\right)  \text{ \ \ \ }\label{HHbar}%
\end{equation}
with $\overline{G}=I$ (so that $\Gamma_{jk}^{i}=0$ and coordinates
$\overline{q}$ are Euclidean) and with the $J_{\overline{G}}$-tensor
$\overline{\mathbf{J}}$ of the form (\ref{og})
\[
\overline{\mathbf{J}}=\left(
\begin{array}
[c]{cc}%
-\overline{q}^{1} & -\frac{1}{2}\overline{q}^{2}\\
-\frac{1}{2}\overline{q}^{2} & 0
\end{array}
\right)  .
\]
The potentials $V$ and $W$ are%
\[
V(\overline{q})=\left(  \overline{q}^{1}\right)  ^{3}+\frac{1}{2}\overline
{q}^{1}\left(  \overline{q}^{2}\right)  ^{2},\text{ \ \ \ }W(\overline
{q})=\frac{1}{4}\left(  \overline{q}^{1}\overline{q}^{2}\right)  ^{2}+\frac
{1}{16}\left(  \overline{q}^{2}\right)  ^{4}.
\]
The system (\ref{HHbar}) has the quasi-bihamiltonian representation
(\ref{bqhbar})
\begin{equation}
\frac{d}{d\overline{t}}\left(
\begin{array}
[c]{c}%
\overline{q}\\
\overline{p}%
\end{array}
\right)  =\overline{\Pi}_{c}d\overline{H}=\frac{1}{\det(\overline{\mathbf{J}%
})}\overline{\Pi}_{nc}(\overline{\mathbf{J}})d\overline{F},\label{Hbar}%
\end{equation}
with Hamiltonians $\overline{H}_{r}$ of the form (\ref{calkibar}). Explicitly:%
\[
\overline{H}_{1}=\overline{H}=\frac{1}{2}\left(  \overline{p}_{1}\right)
^{2}+\frac{1}{2}\left(  \overline{p}_{2}\right)  ^{2}+V(\overline{q}),\text{
\ \ }\overline{H}_{2}=\overline{F}=\frac{1}{2}\overline{q}^{2}\overline{p}%
_{1}\overline{p}_{2}-\frac{1}{2}\overline{q}^{1}(\overline{p}_{2}%
)^{2}+W(\overline{q}).
\]
The non-canonical Poisson operator $\overline{\Pi}_{nc}$ reads explicitly as%

\[
\overline{\Pi}_{nc}(\overline{\mathbf{J}})=\left(
\begin{array}
[c]{cc}%
0 & \overline{\mathbf{J}}\\
-\overline{\mathbf{J}}^{T} & \overline{\Omega}%
\end{array}
\right)  \ \ \ \text{with }\overline{\Omega}=\left(
\begin{array}
[c]{cc}%
0 & -\frac{1}{2}\overline{p}_{2}\\
\frac{1}{2}\overline{p}_{2} & 0
\end{array}
\right)  .
\]
The system (\ref{Hbar}) separates in variables $(\overline{\lambda}%
,\overline{\mu})$ that can be found from the characteristic equation
(\ref{lb}) and are given by%
\begin{align*}
\overline{q}^{1}  & =-(\overline{\lambda}^{1}+\overline{\lambda}^{2})\text{,
}\ \overline{q}^{2}=2\sqrt{-\overline{\lambda}^{1}\overline{\lambda}^{2}}\\
\overline{p}_{1}  & =-\left(  \frac{\overline{\lambda}^{1}\overline{\mu}%
_{1}-\overline{\lambda}^{2}\overline{\mu}_{2}}{\overline{\lambda}%
^{1}-\overline{\lambda}^{2}}\right)  \text{, \ \ }\overline{p}_{2}%
=\sqrt{-\overline{\lambda}^{1}\overline{\lambda}^{2}}\left(  \frac
{\overline{\mu}_{1}-\overline{\mu}_{2}}{\overline{\lambda}^{1}-\overline
{\lambda}^{2}}\right)
\end{align*}
while the separation curve (\ref{SKbar}) generating Hamiltonians $\overline
{H}_{r}$ is%
\begin{equation}
\overline{H}_{1}\overline{\lambda}-\overline{H}_{2}=\frac{1}{2}\overline
{\lambda}\overline{\mu}^{2}-\overline{\lambda}^{4}.\label{ziuta}%
\end{equation}
Let us now take another, arbitrary $J_{\overline{G}}$-tensor $\mathbf{J}_{3}$.
Since the metric $\overline{G}\ =I$ $\ $\ in variables $(\overline{q}%
^{1},\overline{q}^{2})$ the most general form of $\mathbf{J}_{3}$ is
\ (\ref{og}) which reads now as%
\begin{equation}
\mathbf{J}_{3}=\left(
\begin{array}
[c]{cc}%
m\left(  \overline{q}^{1}\right)  ^{2}+2\beta_{1}\overline{q}^{1}+\gamma^{11}
& m\overline{q}^{1}\overline{q}^{2}+\beta_{1}\overline{q}^{2}+\beta
_{2}\overline{q}^{1}+\gamma^{12}\\
m\overline{q}^{1}\overline{q}^{2}+\beta_{1}\overline{q}^{2}+\beta_{2}%
\overline{q}^{1}+\gamma^{12} & m\left(  \overline{q}^{2}\right)  ^{2}%
+2\beta_{2}\overline{q}^{2}+\gamma^{22}%
\end{array}
\right) \label{J}%
\end{equation}
with $6$ arbitrary constants $m,\beta_{i},\gamma^{ij}=\gamma^{ji}$. Using
Proposition \ref{czast3} (with $\mathbf{J}_{1}=I$ and $\mathbf{J}%
_{2}=\overline{\mathbf{J}}$) we see that in a new independent variable defined by%

\[
dt_{3}=\frac{dt}{\det\left(  \mathbf{J}_{3}\right)  }%
\]
our potential-cofactor system (\ref{HHbar}) attains the bi-cofactor form
(\ref{bicoft3}) with $(\Gamma^{(3)})_{jk}^{i}$ being Christoffel symbols of
the metric $G_{3}$ $=\left(  \det\mathbf{J}_{3}\right)  \mathbf{J}_{3}G$. They
can be obtained from (\ref{Gammy}) which reads now (since $\Gamma_{jk}^{i}=0$)%

\[
(\Gamma^{(3)})_{jk}^{i}=-\frac{1}{2\det\mathbf{J}_{3}}\left(  \delta_{j}%
^{i}\frac{\partial\left(  \det\mathbf{J}_{3}\right)  }{\partial\overline
{q}_{k}}+\delta_{k}^{i}\frac{\partial\left(  \det\mathbf{J}_{3}\right)
}{\partial\overline{q}_{j}}\right)  .
\]
It is important to stress that for all the choices of $\mathbf{J}_{3}$ the
obtained system has on $Q$ exactly the same trajectories as Henon-Heiles
system, only traversed with different speed. Moreover, the metric $G_{3}$ is
of constant curvature since it is geodesically equivalent to the flat metric
$\overline{G}=I$ \cite{Beltrami}\textbf{.}

Among all possible choices of the deforming tensor $\mathbf{J}_{3}$ there is
only one that leads to a new potential-cofactor system, namely $\mathbf{J}%
_{3}=\overline{\mathbf{J}}$. This choice leads to cofactor-potential system
(\ref{tilde}) with $\widetilde{\mathbf{J}}=\overline{\mathbf{J}}^{-1}$ and
with the metric $\widetilde{G}$ $=\left(  \det\overline{\mathbf{J}}\right)
\overline{\mathbf{J}}\,\overline{G}$. The metric $\widetilde{G}$ is flat since
the deforming tensor $\mathbf{J}_{3}=\overline{\mathbf{J}}$ satisfies the
conditions (\ref{wkmac}) and (\ref{wksimple}). Explicitly, we have%
\[
\widetilde{\mathbf{J}}=\overline{\mathbf{J}}^{-1}=\frac{4}{\left(
\widetilde{q}^{2}\right)  ^{2}}\left(
\begin{array}
[c]{cc}%
0 & -\frac{1}{2}\widetilde{q}^{2}\\
-\frac{1}{2}\widetilde{q}^{2} & \widetilde{q}^{1}%
\end{array}
\right)  \text{, \ }\widetilde{G}=\frac{1}{4}\left(  \widetilde{q}^{2}\right)
^{2}\left(
\begin{array}
[c]{cc}%
\widetilde{q}^{1} & \frac{1}{2}\widetilde{q}^{2}\\
\frac{1}{2}\widetilde{q}^{2} & 0
\end{array}
\right)
\]
(where of course $\overline{q}^{i}=\widetilde{q}^{i}$). This system has the
quasi-bihamiltonian form (\ref{bqhtilde}) with the Hamiltonians as in
(\ref{Htilde}) ($\widetilde{H}_{1}=$ $\overline{H}_{2}$ and $\widetilde{H}%
_{2}=$ $\overline{H}_{1}$) where the new momenta $\widetilde{p}$ are related
with the old momenta through the map (\ref{bt}) and read explicitly as%
\[
\overline{p}_{1}=-\widetilde{q}^{1}\widetilde{p}_{1}-\frac{1}{2}\widetilde
{q}^{2}\widetilde{p}_{2}\text{, \ }\overline{p}_{2}=-\frac{1}{2}\widetilde
{q}^{2}\widetilde{p}_{1}\text{.}%
\]
Our new system separates in variables $(\widetilde{\lambda},\widetilde{\mu})$
that can be found from the characteristic equation (\ref{lt}) and are given by%
\begin{align*}
\widetilde{q}^{1}  & =-\left(  \frac{1}{\widetilde{\lambda}^{1}}+\frac
{1}{\widetilde{\lambda}^{2}}\right)  \text{, }\ \ \widetilde{q}^{2}=\frac
{2}{\sqrt{-\widetilde{\lambda}^{1}\widetilde{\lambda}^{2}}}\\
\widetilde{p}_{1}  & =-\frac{\widetilde{\lambda}^{1}\widetilde{\lambda}^{2}%
}{\widetilde{\lambda}^{1}-\widetilde{\lambda}^{2}}\left(  \widetilde{\lambda
}^{1}\widetilde{\mu}_{1}-\widetilde{\lambda}^{2}\widetilde{\mu}_{2}\right)
\text{, \ \ }\widetilde{p}_{2}=-\frac{\sqrt{-\widetilde{\lambda}^{1}%
\widetilde{\lambda}^{2}}}{\widetilde{\lambda}^{1}-\widetilde{\lambda}^{2}%
}\left(  \left(  \widetilde{\lambda}^{1}\right)  ^{2}\widetilde{\mu}%
_{1}-\left(  \widetilde{\lambda}^{2}\right)  ^{2}\widetilde{\mu}_{2}\right)  .
\end{align*}
Our system can be obtained from the separation curve of the form
(\ref{SKtilde}) that explicitly reads as%
\[
\widetilde{H}_{1}\widetilde{\lambda}-\widetilde{H}_{2}=-\frac{1}{2}%
\widetilde{\lambda}^{2}\widetilde{\mu}^{2}+\widetilde{\lambda}^{-3}%
\]
and can also be obtained from the separation curve (\ref{ziuta}) by the
transformation (\ref{relf}). Let us now introduce a new coordinates
$(r^{1},r^{2})$ on $Q$ defined through%
\begin{equation}
\widetilde{q}^{1}=-2\frac{r^{1}}{r^{2}}\text{, \ }\widetilde{q}^{2}=\frac
{4}{r^{2}}\text{, }\label{numerek}%
\end{equation}
(see \textbf{\cite{mar}).} In $(r^{1},r^{2})$ the metric $\widetilde{G}$
attains the antidiagonal form
\begin{equation}
\widetilde{G}(r)=\left(
\begin{array}
[c]{cc}%
0 & 1\\
1 & 0
\end{array}
\right) \label{ad}%
\end{equation}
(so that $(r^{1},r^{2})$ are flat coordinates for $\widetilde{G}$ and
$(\widetilde{\Gamma}(r))_{jk}^{i}=0$) while the $J_{\widetilde{G}}$-tensor
$\widetilde{\mathbf{J}}$ becomes%
\[
\widetilde{\mathbf{J}}(r)=\frac{1}{4}\left(
\begin{array}
[c]{cc}%
r^{1}r^{2} & \left(  r^{1}\right)  ^{2}+4\\
\left(  r^{2}\right)  ^{2} & r^{1}r^{2}%
\end{array}
\right)  .
\]
Our cofactor-potential system (geodesically equivalent to (\ref{HHbar}))
attains in variables $(r^{1},r^{2})$ the flat Newton form%

\begin{equation}
\frac{d^{2}}{d\widetilde{t}^{2}}\left(
\begin{array}
[c]{c}%
\widetilde{r}^{1}\\
\widetilde{r}^{2}%
\end{array}
\right)  =-\left(  \operatorname*{cof}\widetilde{\mathbf{J}}(r)\right)
^{-1}\widetilde{\nabla}\widetilde{V}_{2}=-\widetilde{\nabla}\widetilde{V}%
_{1}=\left(  \frac{2}{r^{2}}\right)  ^{5}\left(
\begin{array}
[c]{c}%
2\left(  \left(  r^{1}\right)  ^{2}+1\right) \\
-r^{1}r^{2}%
\end{array}
\right)  \text{ \ \ \ }\label{new}%
\end{equation}
with potentials
\[
\widetilde{V}_{1}(r)=V(r)=16\frac{\left(  \left(  r^{1}\right)  ^{2}+1\right)
}{\left(  r^{2}\right)  ^{4}},\text{ \ \ }\widetilde{V}_{2}(r)=W(r)=-8\frac
{\left(  \left(  r^{1}\right)  ^{2}+2\right)  r^{1}}{\left(  r^{2}\right)
^{3}},
\]
while the corresponding Hamiltonians are%
\begin{align*}
\widetilde{H}_{1}(r,s)  & =\overline{H}_{2}(r,s)=s_{1}s_{2}+\widetilde{V}%
_{1}(r),\\
\widetilde{H}_{2}(r,s)  & =\overline{H}_{1}(r,s)=\left(  \frac{1}{8}\left(
r^{1}\right)  ^{2}+\frac{1}{2}\right)  s_{1}^{2}-\frac{1}{4}r^{1}r^{2}%
s_{1}s_{2}+\frac{1}{8}\left(  r^{2}\right)  ^{2}s_{2}^{2}+\widetilde{V}%
_{2}(r),
\end{align*}
where the momenta $(s_{1},s_{2})$ are obtained from the point transformation
(\ref{numerek}) and are
\[
\widetilde{p}_{1}=-\frac{2}{r^{2}}s_{1},\text{ \ }\widetilde{p}_{2}%
=2\frac{r^{1}}{\left(  r^{2}\right)  ^{2}}s_{1}-\frac{4}{\left(  r^{2}\right)
^{2}}s_{2}.
\]
Let us finally make another choice of the deforming tensor $\mathbf{J}_{3}$,
namely $m=0,$ $\gamma^{22}=a,$ $\beta^{2}=b\neq0$ and $\beta^{1}=\gamma
^{11}=\gamma^{12}=0$ so that
\[
\mathbf{J}_{3}=\left(
\begin{array}
[c]{cc}%
0 & b\overline{q}^{1}\\
b\overline{q}^{1} & 2b\overline{q}^{2}+a
\end{array}
\right)  ,\ \ G_{3}=\left(  \det\mathbf{J}_{3}\right)  \mathbf{J}_{3}%
\overline{G}=\left(
\begin{array}
[c]{cc}%
0 & -b^{3}(\overline{q}^{1})^{3}\\
-b^{3}(\overline{q}^{1})^{3} & -b^{2}(\overline{q}^{1})^{2}(2b\overline{q}%
^{2}+a)
\end{array}
\right)  .
\]
The metric $G_{3}$ is again flat since the deforming tensor $\mathbf{J}_{3}$
satisfies the conditions (\ref{wkmac}) and (\ref{wksimple}). The respective
$J_{G_{3}}$-tensors for the related bi-cofactor system (\ref{bicoft}) are
\[
\mathbf{J}_{1}=\mathbf{J}_{3}^{-1}=\frac{1}{b\overline{q}^{1}}\left(
\begin{array}
[c]{cc}%
-\frac{2b\overline{q}^{2}+a}{b\overline{q}^{1}} & 1\\
1 & 0
\end{array}
\right)  ,\ \ \ \ \mathbf{J}_{2}=\overline{\mathbf{J}}\mathbf{J}_{3}%
^{-1}=\frac{1}{b\overline{q}^{1}}\left(
\begin{array}
[c]{cc}%
\frac{2b\overline{q}^{2}+a}{b}-\frac{1}{2}\overline{q}^{2} & -\overline{q}%
^{1}\\
\frac{1}{2}\frac{\overline{q}^{2}(2b\overline{q}^{2}+a)}{b\overline{q}^{1}} &
-\frac{1}{2}\overline{q}^{2}%
\end{array}
\right)  .
\]
Let us now perform a parameter-dependent change of variables to the
coordinates $(x^{1},x^{2})$ on $Q$ defined through
\[
x^{1}=\frac{1}{2}\frac{2b\overline{q}^{2}+a}{b^{2}\overline{q}^{1}%
},\ \ \ x^{2}=\frac{1}{b^{2}\overline{q}^{1}}.
\]
In $(x^{1},x^{2})$ the metric $G_{3}$ attains the antidiagonal form
(\ref{ad}), so that $(x^{1},x^{2})$ are flat coordinates for $G_{3}$ and the
$J_{G_{3}}$-tensors $\mathbf{J}$ become%
\[
\mathbf{J}_{1}=-b^{2}\left(
\begin{array}
[c]{cc}%
x^{1}x^{2} & (x^{1})^{2}+\frac{1}{b^{2}}\\
(x^{2})^{2} & x^{1}x^{2}%
\end{array}
\right)  ,~\ \ \ \ \ \ \ \mathbf{J}_{2}=\left(
\begin{array}
[c]{cc}%
\frac{1}{2}x^{1}+\frac{1}{4}ax^{2} & \frac{1}{2}ax^{1}\\
x^{2} & \frac{1}{2}x^{1}+\frac{1}{4}ax^{2}%
\end{array}
\right)  .
\]
Hence, our two-parameter family of flat bi-cofactor systems attains in
variables $(x^{1},x^{2})$ the flat Newton form
\begin{align}
\frac{d^{2}}{dt_{3}^{2}}\left(
\begin{array}
[c]{c}%
x^{1}\\
x^{2}%
\end{array}
\right)   & =-\left(  \left(  \operatorname*{cof}\mathbf{J}_{1}\right)
^{-1}\nabla^{(3)}V\right)  ^{i}=-\left(  \left(  \operatorname*{cof}%
\mathbf{J}_{2}\right)  ^{-1}\nabla^{(3)}W\right)  ^{i}\nonumber\\
& \nonumber\\
& =\left(
\begin{array}
[c]{c}%
\dfrac{1}{8}\dfrac{x^{1}[4(x^{1})^{2}-4ax^{1}x^{2}+a^{2}(x^{2})^{2}]}%
{b^{4}(x^{2})^{5}}+\dfrac{1}{2}\dfrac{4x^{1}+ax^{2}}{b^{6}(x^{2})^{5}}\\
\dfrac{1}{8}\dfrac{4(x^{1})^{2}-4ax^{1}x^{2}+a^{2}(x^{2})^{2}}{b^{4}%
(x^{2})^{4}}+\dfrac{3}{b^{6}(x^{2})^{4}}%
\end{array}
\right) \label{bc}%
\end{align}
with the potentials
\begin{align*}
V  & =\frac{1}{8}\frac{4(x^{1})^{2}-4ax^{1}x^{2}+a^{2}(x^{2})^{2}}{b^{4}%
(x^{2})^{3}}+\frac{1}{b^{6}(x^{2})^{3}},\ \ \ \\
& \ \ \\
W  & =\frac{1}{16^{2}}\frac{(ax^{2}-2x^{1})[4(x^{1})^{2}-4ax^{1}x^{2}%
+a^{2}(x^{2})^{2}]}{b^{4}(x^{2})^{4}}+\frac{1}{16}\frac{ax^{2}-2x^{1}}%
{b^{6}(x^{2})^{4}}.
\end{align*}
All the flat bi-cofactor systems in the 2-parameter family (\ref{bc}) are
geodesically equivalent to both the Henon-Heiles system (\ref{HHbar}) and the
Hamiltonian system (\ref{new}). They also belong to the whole family of such
systems generated by (\ref{J}).

\end{document}